\documentclass[fleqn,usenatbib]{mn2e}
\usepackage{times}
\usepackage{amsmath}
\usepackage{mathtools}
\usepackage{graphicx}
\usepackage{booktabs} 
\usepackage{dcolumn}
\usepackage{fixltx2e} 

\newcolumntype{d}[1]{D{.}{.}{#1}} 
\newcolumntype{e}[1]{D{e}{e}{#1}} 

\title[Bayesian time series analysis of terrestrial impact cratering]{Bayesian time series analysis of terrestrial impact cratering}

\author[C.A.L.\ Bailer-Jones]
{C.A.L.\ Bailer-Jones\thanks{Email: calj@mpia.de}\\
Max-Planck-Institut f\"ur Astronomie, K\"onigstuhl 17, 69117 Heidelberg, Germany}

\begin{document}

\date{{\em Submitted 13 March 2011. Revised 11 April 2011 and 17 May 2011. Accepted 20 May 2011. Minor typos corrected 15 June 2011. Minor correction (erratum) 16 October 2011}}

\maketitle

\label{firstpage}

\begin{abstract} 

Giant impacts by comets and asteroids have probably had an important influence on terrestrial biological evolution.  We know of around 180 high velocity impact craters on the Earth with ages up to 2400\,Myr and diameters up to 300\,km.  Some studies have identified a periodicity in their age distribution, with periods ranging from 13 to 50\,Myr. It has further been claimed that such periods may be causally linked to a periodic motion of the solar system through the Galactic plane.  However, many of these studies suffer from methodological problems, for example misinterpretation of p-values, overestimation of significance in the periodogram or a failure to consider plausible alternative models. Here I develop a Bayesian method for this problem in which impacts are treated as a stochastic phenomenon. Models for the time variation of the impact probability are defined and the evidence for them in the geological record is compared using Bayes factors. This probabilistic approach obviates the need for ad hoc statistics, and also makes explicit use of the age uncertainties.  I find strong evidence for a monotonic decrease in the recorded impact rate going back in time over the past 250\,Myr for craters larger than 5\,km. The same is found for the past 150\,Myr when craters with upper age limits are included. This is consistent with a crater preservation/discovery bias modulating an otherwise constant impact rate.  The set of craters larger than 35\,km (so less affected by erosion and infilling) and younger than 400\,Myr are best explained by a constant impact probability model.  A periodic variation in the cratering rate is strongly disfavoured in all data sets. There is also no evidence for a periodicity superimposed on a constant rate or trend, although this more complex signal would be harder to distinguish.

\end{abstract}


\section{Introduction}\label{sect:introduction}

About 180 terrestrial impact craters are known. The high velocity of the impact means that even relatively small comets or asteroids produce large craters. The meteor which crated the 1.2\,km diameter Barringer crater in Arizona, for example, was probably only 50m across. Since the discovery of evidence for a large impact 65\,Myr ago at the geological boundary between the Cretaceous and Tertiary periods (the K-T boundary) (Alvarez et al.~\citealp{alvarez_etal_1980}) and its implication in the mass extinction event at that time (including the demise of the dinosaurs), it has become clear that bolide impacts have had a significant impact on the evolution of life.

The large impactors are believed to be either asteroids from the main asteroid belt, or comets from the Oort cloud (Shoemaker 1983). The multi-body dynamics involved in putting these on a collision course with the Earth implies that cratering is a random phenomenon, but the rate of impacts is not necessarily constant in time. It has been suggested that gravitational perturbations of the Oort cloud due to the Galactic tide, passages of the solar system near to molecular clouds, or an unseen solar companion, may send large numbers of comets into the inner solar system as a comet shower, increasing the impact rate (Davis et al.\ 1984, Torbett \& Smoluchowski~\citealp{torbett_smo_1984}, Rampino \& Stothers 1984, Napier 1998, Wickramasinghe \& Napier 2008, Gardner et al.\ 2011).
Simple dynamical calculations indicate that the Sun oscillates vertically about the Galactic midplane with a period of 52--74\,Myr, depending on the mass density assumed (Bahcall \& Bahcall 1985, Shuter \& Klatt 1986, Stothers 1998).  In parallel to this, several studies claim to have found evidence for a temporal periodicity in the impact cratering record over the past few hundred million years, with numerous periods ranging from 13 to 50\,Myr having been identified (Alvarez \& Muller 1984, Rampino \& Stothers 1984, Grieve et al.\ 1985, Montanari et al.\ 1998, Napier 1998, Yabushita 2004, Chang \& Moon 2005, Napier 2006).  Some authors make a causal link, suggesting that each midplane crossing increases the impact rate.  It has also been suggested that periodicities in cratering may be associated with alleged periodicities in mass extinctions or in biodiversity variations, although there is little evidence linking any mass extinction apart from the K-T one to a giant impact (Alvarez 2003, Hallam 2004; see Bailer-Jones 2009 for a more general review of extraterrestrial influence on terrestrial climate and biodiversity.)
Other studies of the crater record conclude there to be insufficient evidence for periodicity (e.g.\ Grieve 1991, Grieve \& Pesonen 1996, Yabushita 1996, Jetsu \& Pelt 2000).

While these are a priori reasonable suggestions worthy of further analysis, many of the studies claiming to have identified periods are compromised by problems with their methodology. Typical problems are misinterpreting p-values, overestimating the significance of periodogram peaks, or failing to consider a sufficient set of models.  Possibly because the data comprise only crater ages (with no attached magnitudes like more familiar time series), many studies have developed ad hoc statistics to look for periods, many of which have poorly explored statistical properties.  Identifying ``periods'' is relatively easy -- any time series can be expressed as a sum of Fourier terms; clusters of points can always be found -- but properly assessing significance is harder.  A common mistake is to interpret evidence {\em against} a null hypothesis of ``random data'' as evidence {\em for} some periodic model, neglecting to realise that both may be inferior to a plausible third alternative. Although these are known limitations of frequentist hypothesis testing which have been discussed extensively (e.g.\ Berger \& Sellke 1987, Kass \& Raftery 1996, Marden 2000, Berger~\citealp{berger_2003}, Jaynes~\citealp{jaynes_2003}, MacKay 2003, Christensen~\citealp{christensen_2005}), this seems not to have discouraged their use.  

The aim of this article is to analyse the crater time series with well-motivated statistical methods. One of the key features is to write down explicit models for the impact phenomenon. A second feature is that I consider this phenomenon to be a {\em stochastic} process: Rather than expecting the impact events to follow a deterministic pattern, I model the time variation of the impact probability. This better accommodates the astrophysical and geological contexts 
(e.g.\ smooth variations in the torques on the Oort cloud, or slow erosion of craters).
By using the Bayesian framework to analyse the data, we can properly calculate the evidence for the different models and compare them on an equal footing.  
The critical aspect is that we must compare evidence for the entire model (the average likelihood over the model parameter space), rather than comparing the tuned maximum likelihood fit (which generally favours the more complex model).

Craters are difficult to date, and some have very large age uncertainties (Grieve 1991, Deutsch \& Sch\"arer 1994).  There has been a lot of agonizing in the literature about how to deal with age uncertainties, and some studies may have biased their conclusions by removing ``poor'' data. Contrary to claims in the literature, including craters with large uncertainties is not a problem for model comparison {\em provided} we use these uncertainties appropriately.  The method developed in this paper does this, and it can also include craters which only have upper age limits.

The impact crater record is certainly incomplete. Of most relevance is the preservation bias: on account of erosion and infilling, older, smaller craters are less likely to survive or to be discovered. The oldest known crater is about 2400\,Myr old, but there is an obvious paucity of craters older than about 700\,Myr (only 14 of 176 older than this).  For this reason I focus my analysis on craters larger than 5\,km in diameter which have ages (or upper age limits) below 250\,Myr, although I also extend the analysis back to 400\,Myr BP (before present). Rather than attempting to ``de-bias'' the data, I model the data as they are, so strictly I am modelling not the original impact history (which we do not observe), but the combined impact/preservation history.

After outlining the data in section~\ref{sect:data}, I describe the method in section~\ref{sect:method}.
This is tested and demonstrated on simulated data in section~\ref{sect:simulations}.
The results are presented in section~\ref{sect:results}, followed by a discussion of these and the method and a comparison with other studies in section~\ref{sect:discussion}. I summarize and conclude in section~\ref{sect:conclusions}.
I will say very little about the wider topics of the comet and asteroid population, impact effects, crater geology and aging etc. These are reviewed by Shoemaker (1983), Grieve (1991), Deutsch \& Sch\"arer (1994) and Grieve \& Pesonen (1996), amongst others.

\section{Impact crater data}\label{sect:data}

\begin{table}[h]
\begin{center}
\caption{The 59 craters in the Earth Impact Database with diameters greater than or equal to 5\,km and ages or age upper limits below 250\,Myr
\label{eidmain}}
\vspace*{1em}
\begin{tabular}{l d{2} d{3} d{1} }
\toprule
Name & \multicolumn{1}{c}{{\rm age}} & \multicolumn{1}{c}{{\rm $\sigma$(age)}} & \multicolumn{1}{c}{{\rm diameter}} \\
  & \multicolumn{1}{c}{{\rm Myr}} & \multicolumn{1}{c}{{\rm Myr}} & \multicolumn{1}{c}{{\rm km}} \\
\midrule
           Araguainha &  244.4 &  3.25 &    40 \\
                 Avak &     49.0 &     28.0 &    12 \\
  Beyenchime-Salaatin &     40 &    20 &     8 \\
               Bigach &      5 &     3 &     8 \\
              Boltysh &  65.17 &  0.64 &    24 \\
             Bosumtwi &   1.07 &     0.107 &  10.5 \\
             Carswell &    115 &    10 &    39 \\
       Chesapeake Bay &   35.3 &   0.1 &    90 \\
            Chicxulub &  64.98 &  0.05 &   170 \\
               Chiyli &     46 &     7 &   5.5 \\
              Chukcha &     <70 &      &     6 \\
          Cloud Creek &    190 &    30 &     7 \\
       Connolly Basin &     <60 &      &     9 \\
             Deep Bay &     99 &     4 &    13 \\
               Dellen &     89 &   2.7 &    19 \\
          Eagle Butte &     <65 &      &    10 \\
          El'gygytgyn &    3.5 &   0.5 &    18 \\
         Goat Paddock &     <50 &      &   5.1 \\
         Gosses Bluff &  142.5 &   0.8 &    22 \\
             Haughton &     39 &     3.9 &    23 \\
 Jebel Waqf as Suwwan &    46.5 &    5.8 &   5.5 \\
              Kamensk &     49 &   0.2 &    25 \\
                 Kara &   70.3 &   2.2 &    65 \\
             Kara-Kul &      <5 &      &    52 \\
                Karla &      5 &     1 &    10 \\
             Kentland &     <97 &      &    13 \\
                Kursk &    250 &    80 &     6 \\
           Lappajärvi &   73.3 &   5.3 &    23 \\
             Logancha &     40 &    20 &    20 \\
              Logoisk &   42.3 &   1.1 &    15 \\
          Manicouagan &    214 &     1 &   100 \\
               Manson &   74.1 &   0.1 &    35 \\
          Maple Creek &     <75 &      &     6 \\
              Marquez &     58 &     2 &  12.7 \\
                 Mien &    121 &   2.3 &     9 \\
            Mistastin &   36.4 &     4 &    28 \\
              Mjølnir &    142 &   2.6 &    40 \\
           Montagnais &   50.5 &  0.76 &    45 \\
            Morokweng &    145 &   0.8 &    70 \\
                Oasis &    <120 &      &    18 \\
              Obolon' &    169 &     7 &    20 \\
              Popigai &   35.7 &   0.2 &   100 \\
      Puchezh-Katunki &    167 &     3 &    80 \\
            Ragozinka &     46 &     3 &     9 \\
             Red Wing &    200 &    25 &     9 \\
                 Ries &   15.1 &   0.1 &    24 \\
         Rochechouart &    214 &     8 &    23 \\
         Saint Martin &    220 &    32 &    40 \\
        Sierra Madera &    <100 &      &    13 \\
          Steen River &     91 &     7 &    25 \\
            Tin Bider &     <70 &      &     6 \\
          Tookoonooka &    128 &     5 &    55 \\
        Upheaval Dome &    <170 &     &    10 \\
         Vargeao Dome &     <70 &      &    12 \\
         Vista Alegre &     <65 &     &   9.5 \\
            Wanapitei &   37.2 &   1.2 &   7.5 \\
          Wells Creek &    200 &   100 &    12 \\
             Wetumpka &     81 &   1.5 &   6.5 \\
           Zhamanshin &    0.9 &   0.1 &    14 \\
\bottomrule
\end{tabular}
\end{center}
\end{table}

The data are taken from the {\em Earth Impact Database} (EID), a compilation from the literature maintained by the Planetary and Space Science Centre at the University of New Brunswick.\footnote{http://www.passc.net/EarthImpactDatabase/}  This has been the source of data for many previously published studies, and has been continuously expanded as new craters have been discovered, and information revised as improved age or size estimates obtained.  
As of 30 September 2010 this listed 176 craters.
%
My study focuses primarily on craters younger than 250\,Myr with diameters $d>5$\,km. 59 craters fulfil these criteria and are listed in Table~\ref{eidmain}.  
42 of these have an age and age uncertainty in the EID. These uncertainties are interpreted as 1$\sigma$ Gaussian uncertainties (see section~\ref{sect:measmod} for why this is so).
The data on the remaining 17 craters fall into three groups:
\begin{itemize}

\item 13 craters have only upper limits to their ages. These can be included in the analysis consistently, as outlined in section~\ref{sect:censored}.

\item Two craters, {\em Bosumtwi} and {\em Haughton}, have no age uncertainty reported in the EID. Craters younger than 50\,Myr have dating errors ranging from 0.1 to 20\,Myr (or 0.3\% to 60\%), so it is difficult to estimate an appropriate uncertainty. I rather arbitrarily assign 10\% uncertainties to the ages of these two craters.

\item Two craters, {\em Avak} and {\em Jebel Waqf as Suwwan}, have age ranges (3--95\,Myr and 37--56\,Myr respectively) rather than estimates with uncertainties in the EID.  I consider the range as a 90\% confidence interval of a Gaussian distribution with mean equal to the average of the limits (so the standard deviation is 0.30 times the range). 

\end{itemize}
Crater diameters are notoriously difficult to measure. No uncertainties are listed in the EID, but the uncertainty can be a factor of two or more for buried craters (e.g.\ Grieve 1991).  On account of erosion and infilling, the older or smaller a crater is, the less likely it is to be preserved or discovered, resulting in increasing incompleteness with look back time.  For this reason -- and following several other studies -- I limit my analysis to craters larger than 5\,km in diameter.  Our knowledge of geological processes suggests that for times back to 150--250\,Myr, most craters larger than 5\,km should be reasonably well preserved (although this is something I will test).  Note that I only use the diameters to select the sample; they are not used in the actual analysis. 

We certainly have not identified all impacts: the continents have not been explored equally thoroughly, and craters from marine impacts are rarer and harder to identify. But provided these selection effects are time independent (and size independent above 5\,km), they do not introduce any relevant biases.

The continents have also moved. 250\,Myr, ago at the Permian--Triassic boundary, there was a larger concentration of land mass towards southern latitudes than there is now. Given that asteroid impactors are not distributed isotropically -- their orbits are concentrated in the ecliptic -- we may expect this continental drift to introduce a time variation in the impact rate. However, calculations by Le Feuvre \& Wieczorek (2008) show that the impact probability actually has only a very weak latitudinal dependence, being only 4\% lower at the poles than at the equator.  The net dependence will be even lower when (isotropic) comets are included. Thus continental drift produces a very minor bias which I ignore.

\begin{figure*} 
\begin{center}\includegraphics[width=0.9\textwidth, angle=0]{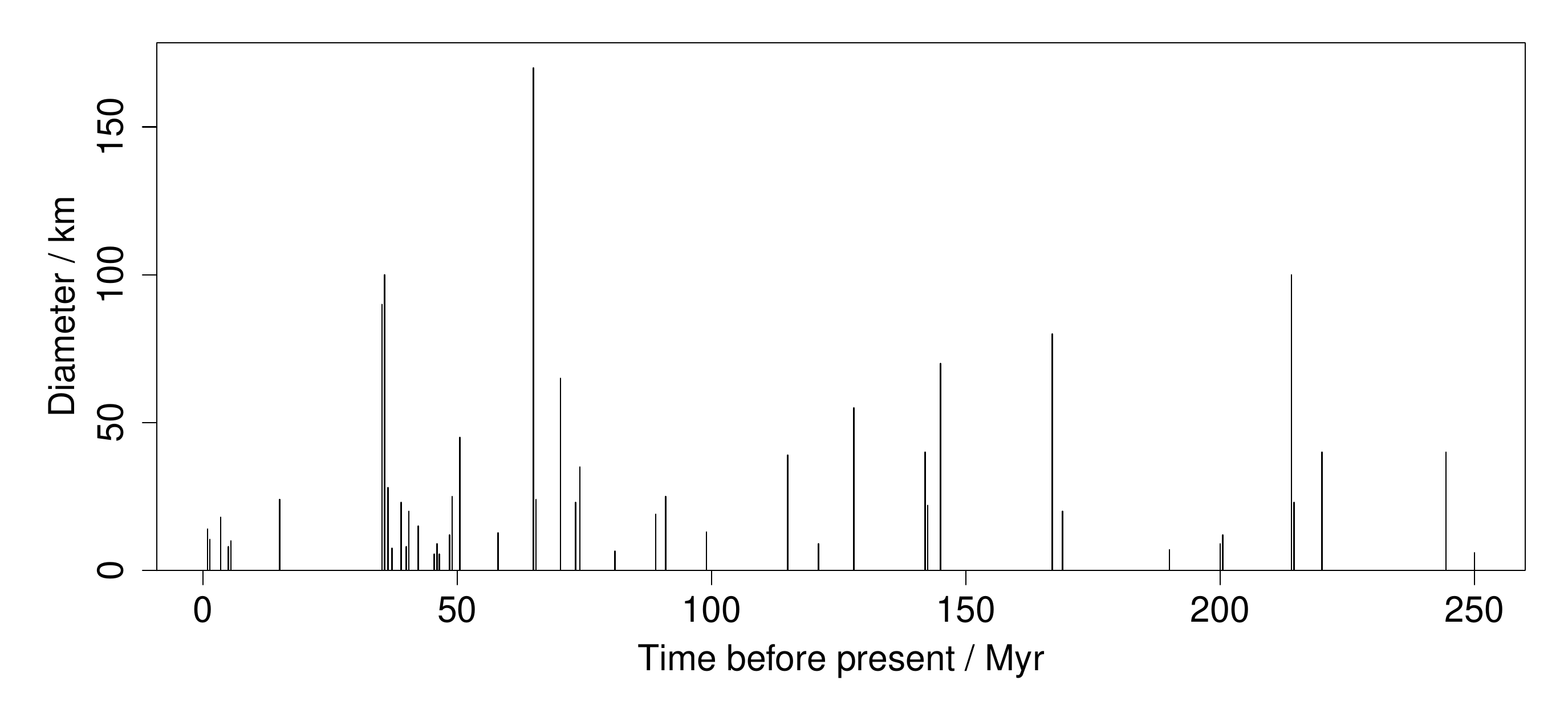}
\caption{The 59 craters listed in Table~\ref{eidmain}, excluding the 13 crater with upper limits on their ages. Several craters have identical or very similar ages so for the sake of this plot I shifted some ages by up to 0.5\,Myr in order to distinguish them.
\label{fig:agediam_1}}
\end{center}
\end{figure*}

\begin{figure*} 
\begin{center}
\includegraphics[width=0.9\textwidth, angle=0]{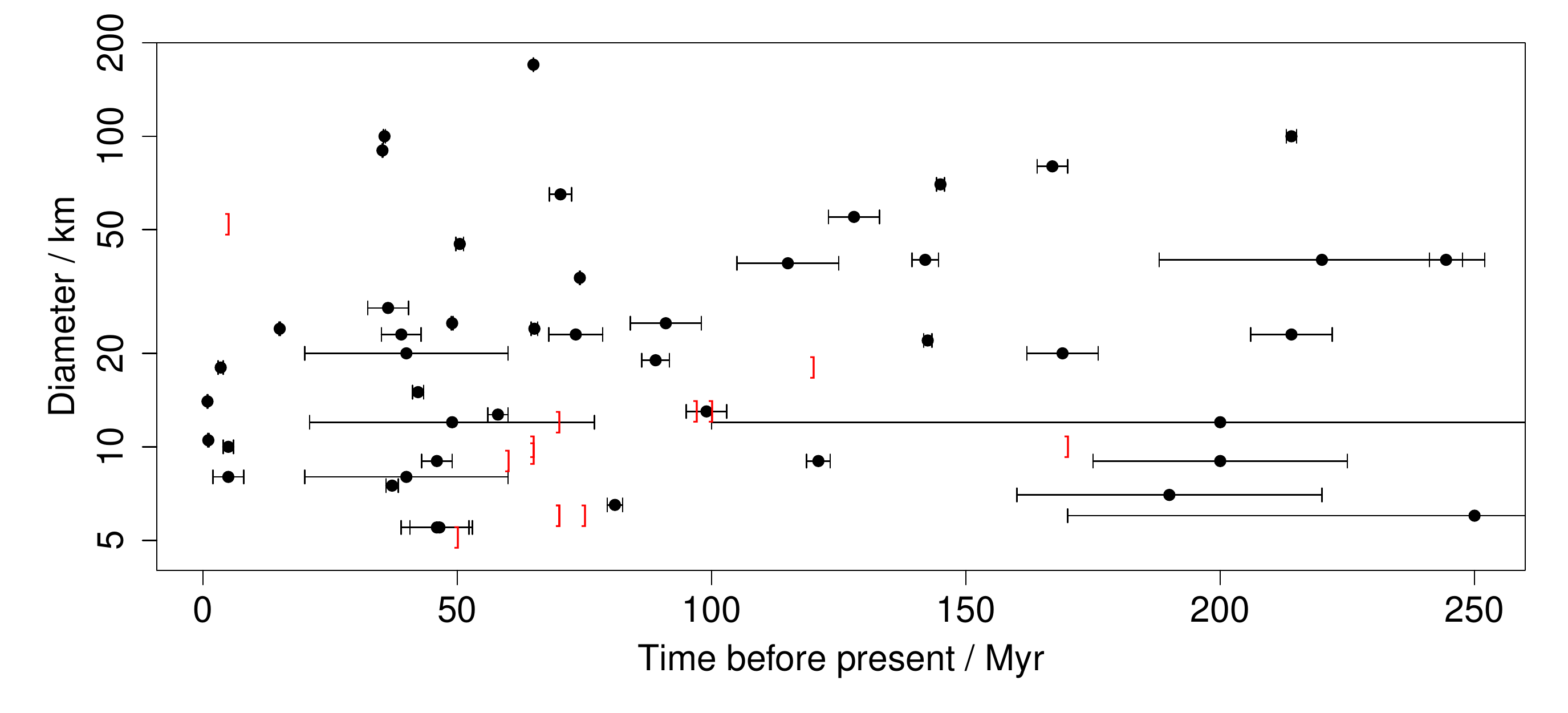}
\caption{The 59 craters listed in Table~\ref{eidmain}, including the 13 craters with upper limit ages (plotted with the ``]'' symbol). There are two upper identical points at 70\,Myr, 6\,km. The error bars are the age uncertainties and in some cases are smaller than the size of the points. Note that diameter is plotted on a logarithmic scale.
\label{fig:agediam_2}}
\end{center}
\end{figure*}

Fig.~\ref{fig:agediam_1} shows the ages and diameters of the 46 craters in Table~\ref{eidmain} which have age estimates (rather than upper limits). Fig.~\ref{fig:agediam_2} show the ages uncertainties on these points and includes the 13 craters with upper limits on their ages. 

I perform the analysis on several different data sets over three ages ranges.
Craters with age estimates and uncertainties ($\sigma_t$) originating from the EID form the {\em basic} data sets. Adding to these the craters for which I have assigned ages/age uncertainties forms the {\em extended} data sets. Further adding craters with upper age limits ($s^{\rm up}$) gives the {\em full} data sets.  I do not make any cut on the age uncertainties. The data sets are as follows
\begin{tabbing}
\= {\bf basic150} \= (32 craters) \= age\,$\leq\,150$\,Myr, $\sigma_t$ original \\
\> {\bf ext150} \> (36 craters) \= age\,$\leq\,150$\,Myr, $\sigma_t$ original or assigned \\
\> {\bf full150} \> (48 craters) \= ext150 plus craters with $s^{\rm up}\leq\,150$\,Myr \\
\> {\bf basic250} \> (42 craters) \= age\,$\leq\,250$\,Myr, $\sigma_t$ original \\
\> {\bf ext250} \> (46 craters) \= age\,$\leq\,250$\,Myr, $\sigma_t$ original or assigned \\
\> {\bf full250} \> (59 craters) \= ext250 plus craters with $s^{\rm up}\leq\,250$\,Myr \\
\> {\bf large400} \> (18 craters) \= age\,$\leq\,400$\,Myr, $d>$\,35\,km, $\sigma_t$ original
\end{tabbing}
This final data set, large400, extends further back in time, but now we definitely expect a bias of preferential preservation and discovery for larger craters above 5\,km. Following previous studies (see section~\ref{sect:discussion}), I therefore just retain craters with much large diameters in an attempt to avoid this bias.  In addition to the 14 craters matching craters from Table~\ref{eidmain}, the four additional (older) craters in this set are: Clearwater West ($290 \pm 20$\,Myr, 36\,km); Charlevoix ($342 \pm 15$\,Myr, 54\,km); Woodleigh ($364 \pm 8$\,Myr, 40\,km); Siljan ($376.8 \pm 1.7$\,Myr, 52\,km).

\section{Method}\label{sect:method}

I now introduce the time series analysis method, showing how to model a generic time-of-arrival data set with a probabilistic model and how to calculate its evidence.

\subsection{Bayesian hypothesis testing}

The goal of hypothesis testing is to identify which of a set of hypotheses is best supported by the data.  More quantitatively, we would like to determine $P(M|D)$, the probability that a hypothesis or model $M$ is true given a set of measured data $D$. Here $D$ is the ages for a set of craters. $M$ could be ``uniform distribution'' or ``periodic distribution'', for example.

Perhaps surprisingly, orthodox (or frequentist) statistics lacks a general framework for this problem and offers instead a number of recipes based on defining some statistic. These normally involve calculating the value for that statistic (e.g.\ $\chi^2$), and comparing it with the value which would be achieved by some ``random'' (noise) model. As discussed at some length in the literature, some of these techniques are inconsistent or misleading, even when we just have two alternative hypotheses
(e.g.\ Berger \& Sellke 1987, Kass \& Raftery 1996, Berger~\citealp{berger_2003}, Christensen~\citealp{christensen_2005},
Bailer-Jones~\citealp{cbj09}; see also section~\ref{sect:discussion}).
The Bayesian approach, in contrast, is direct and often turns out to be quite simple. It inevitably involves a number of numerical integrals, but these can be solved with computers. For more background on Bayesian techniques in general see Jeffreys (2000), Jaynes~\citep{jaynes_2003}, MacKay (2003) or Gregory (2005).

To calculate $P(M|D)$ for one particular model $M_0$ we use Bayes' theorem
\begin{eqnarray}
P(M_0|D) \, &=& \, \frac{P(D|M_0)P(M_0)}{P(D)} \nonumber \\
                  &=& \, \frac{P(D|M_0)P(M_0)}{\sum\limits_{k=0}^{k=K} P(D|M_k)P(M_k)} \nonumber \\
                  &=& \, \frac{1}{1 + \frac{ \sum_{k=1}^{k=K}  P(D|M_k)P(M_k)}{P(D|M_0)P(M_0)}  } 
\label{eqn:modelpost}
\end{eqnarray}
where $k = 0 \ldots K$ represents all plausible models. If there are only two, $M_0$ and $M_1$, this simplifies to
\begin{equation}
P(M_0 | D) \,=\,  \left[ 1 + \frac{ P(D|M_1)P(M_1)}{P(D|M_0)P(M_0)} \right]^{-1} \ .
\label{eqn:modelpost2}
\end{equation}
This follows because implausible models are -- by definition -- those with negligible model prior probabilities, $P(M) \ll 1$.
$P(D|M)$ is called the {\em evidence} for model $M$ (derived in the next section). If we assign the two models equal prior probabilities, then the evidence ratio alone determines the posterior probability, $P(M_0 | D)$.
This evidence ratio is called the {\em Bayes factor}
\begin{equation}
BF_{10} \,=\, \frac{P(D|M_1)}{P(D | M_0)}
\label{eqn:bayesfactor}
\end{equation}
of model 1 with respect to model 0.
When $BF_{10}=1$ the posterior probability is 0.5 for both models.
When $BF_{10}\gg1$ then $P(M_0 | D) \simeq 1/BF$, and when $BF_{10}\ll1$ then $P(M_0 | D) \simeq 1-BF_{10}$. 
If we calculate Bayes factors greater than 10 or less than 0.1 then we can start to claim ``significant'' evidence for one model over the other
(e.g.\  Kass \& Raftery 1996, Jeffreys 2000).
I shall use Bayes factors throughout this article to compare models. 

Given the Bayes factors for all models relative to $M_0$, the posterior probability of this model is then
\begin{equation}
P(M_0 | D) \,=\,  \left[ 1 + \sum_{k=1}^{k=K} BF_{k0} \frac{P(M_k)}{P(M_0)} \right]^{-1} \ .
\label{eqn:modelpost3}
\end{equation}
One difficulty of the Bayesian approach is
that in order to calculate this posterior probability one must specify all plausible models (in order to get the correct summation needed to normalize the probabilities).  This is often not possible (other than for simple two-way hypotheses).
Yet even when we cannot identify all models, Bayes factors remain a valid way of comparing the relative merits of a set of models, and thus identifying the best of these.

\subsection{Measurement model for time-of-arrival data}\label{sect:measmod}

The critical characteristic of the crater time series is that it is just a list of times (with uncertainties), without any corresponding quantity. 
This is unlike most other time series encountered in astrophysics, such as a light curves or radial velocity time series. 
(Some authors have used crater size or inferred impact energy as the ``dependent variable'' in a time series analysis (e.g.\ Yabushita 2004), but this is risky given the significant uncertainties in diameters.)  

Consider a single event $j$, with measured age $s_j$ and corresponding uncertainty $\sigma_j$. This measured age is an {\em estimate} of the (unknown) true age, $t_j$. We express this uncertainty probabilistically. Assuming
a Gaussian distribution for the measurement, the probability of observing the measurement 
$s_j$ given the true age is
\begin{equation}
P(s_j | \sigma_j, t_j) = \frac{1}{\sqrt{2\pi}\sigma_j} \, e^{-(s_j - t_j)^2/2\sigma_j^2} 
\label{eqn:gaussage}
\end{equation}
which is normalized with respect to $s_j$.\footnote{The maximum entropy principle tells us that if we only have the mean and standard deviation of a quantity, then the least informative (most conservative) probability distribution for this quantity is the Gaussian. Of course, if the standard deviation is a significant fraction of the age, then a Gaussian has a significant probabilty mass at negative age. As this problem effects no more than five events listed in the table, I consider this as an adequate approximation.}

\subsection{Stochastic time series models}

The goal is to compare plausible models which could produce the observed impact crater time series. Given the astrophysical context, we do not expect the sequence of crater impacts to be deterministic. For example, when we postulate that the time series is ``periodic'', we do not expect the events to have a strict spacing, even if the ages were measured arbitrarily accurately. An exact, intrinsic rhythm perturbed only by measurement errors seems highly implausible a priori. We should instead understand periodic to mean a periodically varying probability of an impact.
This is a {\em stochastic} model. It is described by
$P(t_j | \theta, M)$, the probability of getting an impact at time $t_j$ given model $M$ with parameters $\theta$. 
A simple periodic model would be a sinusoid, described by the two parameters period and phase. 

\subsection{Bayesian evidence for time-of-arrival data}\label{sect:derivation}

\begin{figure} 
\begin{center}
\includegraphics[width=0.40\textwidth, angle=0]{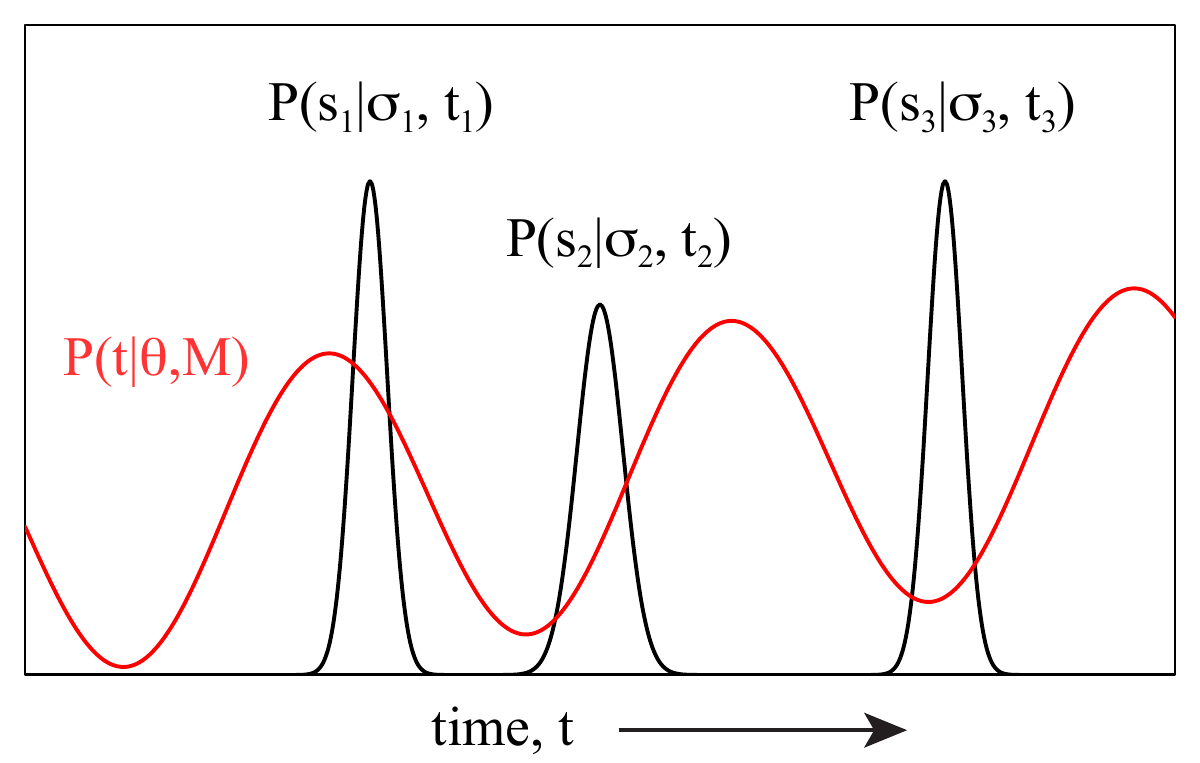}
\caption{Principle of the likelihood calculation (equation~\ref{eqn:likelihood})
\label{fig:likecalcprinciple}}
\end{center}
\end{figure}

I now put the above considerations together to derive the Bayesian evidence, $P(D|M)$.

The probability of observing data $s_j$ from model $M$ with parameters $\theta$ is 
$P(s_j | \sigma_j, \theta, M)$, the {\em likelihood} for one event. The time series model predicts the true age of an event, which is unknown. Applying the rules of probability we marginalize over this to get
\begin{eqnarray}
P(s_j | \sigma_j, \theta, M) &\,=\,& \int_{t_j} P(s_j, t_j | \sigma_j, \theta, M) dt_j \nonumber \\ 
                            &\,=\,& \int_{t_j} P(s_j | \sigma_j, t_j, \theta, M) P(t_j | \sigma_j, \theta, M) dt_j \nonumber \\
                            &\,=\,& \int_{t_j} P(s_j | \sigma_j, t_j) P(t_j | \theta, M) dt_j \ .
\label{eqn:likederiv}
\end{eqnarray}
The last step follows from conditional independence: in the first term -- the measurement model --  once $t_j$ is specified $s_j$ becomes conditionally independent of $\theta$ and $M$; in the second term -- the time series model -- $t_j$ is independent of $\sigma_j$.
As the data are fixed, we consider both terms as functions of $t_j$. Both must be properly normalized probability density functions.

If we have a set of $J$ events for which the ages and uncertainties have been estimated independently of one another, then the probability of 
observing these data $D=\{s_j\}$, the {\em likelihood}, is
\begin{eqnarray}
P(D | \sigma, \theta, M) &\,=\,& \prod_j P(s_j | \sigma_j, \theta, M) \nonumber \\
                         &\,=\,& \prod_j \int_{t_j} P(s_j | \sigma_j, t_j) P(t_j | \theta, M) dt_j \ .
\label{eqn:likelihood}
\end{eqnarray}
where $\sigma = \{\sigma_j\}$.
The principle of this calculation is illustrated in Fig.~\ref{fig:likecalcprinciple}: the likelihood of an event for a given model is the integral of the probability distribution for the event (eqn.~\ref{eqn:gaussage}) over the time series model, $P(t_j | \theta, M)$. Specific cases for the latter are introduced in section~3.6 below.

The {\em evidence} is obtained by marginalizing the likelihood over the parameter prior probability distribution, $P(\theta | M)$,
\begin{eqnarray}
P(D | \sigma, M)  &\,=\,&  \int_{\theta} P(D, \theta | \sigma,  M) d\theta \nonumber \\ 
                           &\,=\,&  \int_{\theta} P(D | \sigma, \theta, M) P(\theta | M) d\theta  
\label{eqn:evidence}
\end{eqnarray}
(where $\sigma$ drops out of the second term due to conditional independence).
For a given set of data (crater time series), we calculate this evidence for the different models we wish to compare, each parametrized by some parameters $\theta$.  The parameter prior $P(\theta | M)$ 
encapsulates our prior knowledge (i.e.\ independent of the data) of the probabilities of different parameters, normally established from the context of the problem (see section~3.7). Note that the evidence is dependent on the measurement uncertainties (although I drop this explicit conditioning in the rest of this article).

A fundamental aspect of this evidence framework for model assessment is that we are not interested in the ``optimal'' value of the parameters for some model.  We are not interested in ``fitting'' the model, but rather in determining how good the model is overall at explaining the data.  The evidence measures this by averaging the likelihood, which is defined at fixed $\theta$, over the prior for $\theta$.  The prior probability distribution is therefore important, and should be considered as part of the model definition.  For example, a periodic model with a non-zero prior over the period range 10--50\,Myr is distinct from a model with a non-zero prior over the range 50--100\,Myr.  The reason why we should integrate over the possible solutions rather than selecting the best one will be illustrated and discussed in some detail.

\subsection{Inclusion of events with age limits}\label{sect:censored}


By representing the unknown, true age of an event probabilistically, we take into account the age uncertainties. It also allows us to include events which only have upper or lower limits on their ages (censored data). 


Given a (hard) upper limit, $s^{\rm up}_j$, all we know is that the true age is younger.  A suitable -- and uninformative -- measurement model is to assume that the probability of measuring some value for the upper limit is constant for any age below the one we actually measured, $s^{\rm up}_j$, and zero otherwise.  In this case, we write equation~\ref{eqn:gaussage} as \begin{equation}
  P(s^{\rm up}_j | t_j) =  \begin{dcases}
  \frac{1}{s^{\rm up}_j}  & \:{\rm when}~~ 0  < t_j < s^{\rm up}_j \\
  0  & \:{\rm otherwise}
\end{dcases}
\end{equation}
which is normalized. 
Recall that time increases into the past.

Events which have lower limits to their ages may be treated in the same way, provided we can also assign an upper limit to the possible age (required to normalize the probability density function). It is not obvious how to assign this. We might use the oldest event in the data set or the age of the oldest rocks searched for craters. As only two events with lower limits occur in the time ranged analysed (and both quite low, at 35\,Myr), these are not included.

\subsection{Impact cratering time series models}\label{sect:tsmodels}

\begin{table}
\begin{center}
\caption{Stochastic time series models.Time $t$ increases into the past. {\em SigProb:Neg} and {\em SigProb:Pos} are special cases of {\em SigProb} with $\lambda<0$ and $\lambda>0$ respectively. {\em SinProbX:Y} specifies {\em SinProb} for $X<T<Y$ (in Myr). \label{tab:tsmodels}}\vspace*{1em}
\begin{tabular}{l l p{2cm}}
\toprule
Name & $P_u(t | \theta, M)$ & \mbox{parameters} \\  \midrule
UniProb  &  1  & none  \\
SinProb & $\frac{1}{2}\left(\cos[2\pi( t/T + \phi)] + 1\right)$  & $T$, $\phi$  \\
SinBkgProb & $\frac{1}{2}\left(\cos[2\pi( t/T + \phi)] + 1\right) + b$   &  $T$, $\phi$, $b$ \\
SigProb  &  $ \left(1+ e^{-(t - t_0)/\lambda} \right)^{-1}$  &  $\lambda$, $t_0$  \\
SinSigProb & SinProb + SigProb & $T$, $\phi$, $\lambda$, $t_0$ \\
\bottomrule
\end{tabular}
\newline
\end{center}
\end{table}

Table~\ref{tab:tsmodels} lists the time series models which will be tested.
The second column gives the (unnormalized) probability density function that an event occurs at time $t$ for given parameters. 
(For given parameters, these distributions must be normalized over the time span of the data when they are used in the likelihood integral.)  {\em SinProb} is a sinusoid with period $T$ and phase $\phi$. {\em SinBkgProb} adds a constant background, $b$, to this. {\em SigProb} is a sigmoidal function parametrized by the steepness of the slope ($\lambda$) and the time of the centre of the slope ($t_0$). It is used to model a monotonic trend in the event probability, with $\lambda<0$ giving a decrease in probability with (look back) time.  {\em SinSigProb} models a a periodic signal on top of a trend.  Examples of these functions are shown later.

\subsection{Choice of parametrization and parameter prior distribution}\label{sect:priorchoice}

To calculate the evidence (equation~\ref{eqn:evidence}) for the models, we integrate over the parameters. In order that this integral converge we must either adopt a proper prior or, equivalently, we must adopt a finite parameter range over which the model is defined. In the interests of keeping assumptions limited and the results intuitive, I adopt a uniform prior over a finite range for all parameters,
\begin{equation}
P_u(\theta | M)  =  \begin{dcases}
  1  & \:{\rm when}~~ \theta_{\rm min}  < \theta < \theta_{\rm max} \\
  0  & \:{\rm otherwise}
\end{dcases}
\end{equation}
The $u$ subscript denotes that this is not normalized (divide by $\Delta \theta= \theta_{\rm max} - \theta_{\rm min} $ to normalize)\footnote{Normalization is essential. It ensures that model complexity is taken into account by the model comparison. Note, therefore, that we are not modelling the absolute rate of impacts, just the probability.}.

It is important to realise that the adopted parameter range is an intrinsic part of the model.  Hence, rather than talking about the model {\em SinProb}, for example, we should talk about the model defined over some period range. For this reason I will refer to models like {\em SinProb10:50}, which means the {\em SinProb} model for the period range 10--50\,Myr.

The evidence of course also depends on the shape of the prior, and I adopt a uniform prior mostly to represent ignorance. But uniform over what? It would be equally valid to parametrize the periodic models in terms of frequency ($\omega=1/T$) and adopt a uniform prior over that, for example. 
The transformation between the priors is $P_T = P_{\omega} T^{-2}$, so a prior uniform in frequency is non-uniform in period. In particular, shorter periods will achieve more weight in the evidence calculation. To assess the impact of this choice, I have repeated all of the analyses
described in sections~\ref{sect:simulations} (simulations) and~\ref{sect:results} (EID results)
which use periodic models with a prior uniform in frequency. It turns out that this change makes bi significant difference to the results, and does not change the conclusions. (The issue of priors is discussed further in section~\ref{sect:discussion}.)

The ranges for the other parameters are as follows. The phase parameter is only defined in the range 0--1, so it is natural to always use this full range (no reason to exclude any phases). For the background parameter $b$ in {\em SinBkgProb}, I examine the evidence over a range 0--5, the upper limit set by the intuition that with much larger backgrounds it will be hard to detect a periodic signal (this was later gound to be the case). For the parameters of {\em SigProb}, I cover the range of $\lambda$ from 0 to $\pm100$. This encompasses models ranging from a step function ($\lambda=0$) to a virtually flat function over the period of interest (see the figures in section~\ref{sect:simulations}). The range of $t_0$ is chosen as 0--275 in order to move the ``crossing point'' of the sigmoid across the whole basic250 time range.

\subsection{Numerical estimation}

The integral in equation~\ref{eqn:evidence} is estimated numerically using
\begin{equation}
\int_x f(x) dx \:\approx\: \sum_{n=1}^{n=N} f(x_n) \delta x \:=\: \frac{\Delta X}{N} \sum_{n=1}^{n=N} f(x_n)
\label{intapprox}
\end{equation}
where $f(x)$ is any continuous integrable function,
the sample $\{x_n\}$ is drawn from a uniform distribution, $\Delta X$ is the range of $x$ from which they are drawn and 
$\delta x = \Delta X / N$ is the average spacing between the samples. 
(For equation~\ref{eqn:likelihood} I just used a regular dense sampling.)
$P(\theta | M) = P_u(\theta | M) / \Delta\theta = 1/\Delta\theta$ for all the priors I use, so
the numerical integration simplifies to
\begin{equation}
P(D | M) \, \approx \, \frac{1}{N}  \sum_{\theta=\theta_{\rm min}}^{\theta=\theta_{\rm max}} P(D | \theta, M)
\end{equation}
with the samples drawn from a random uniform distribution.
Thus with uniform priors, the evidence is just the likelihood averaged over the range of the parameters.
This is evaluated at several million randomly selected points, more than enough to ensure a very high signal-to-noise ratio in the calculated evidence.


\section{Tests on simulated time series} \label{sect:simulations}

To test the method I apply it to numerous simulated data sets.  The goal is to determine whether we can identify the underlying signature in the data, and with what significance, and whether we sometimes identify the wrong model. I will demonstrate, for example, that simply identifying a ``significant'' peak in the periodogram can lead us to the wrong conclusion.

I generate stochastic time series using the same probabilistic models described in section~\ref{sect:tsmodels}.  For each model and set of parameters (e.g.\ period and phase in {\em SinProb}) I draw events independently at random from the probability distribution.
I generate several random time series for each parameter combination.  In all cases a time series comprises 42 events over a time span of 250\,Myr -- characteristic of the basic250 data set -- and $\sigma_t$ is unity for all events.

\begin{figure} 
\begin{center}
\includegraphics[width=0.50\textwidth, angle=0]{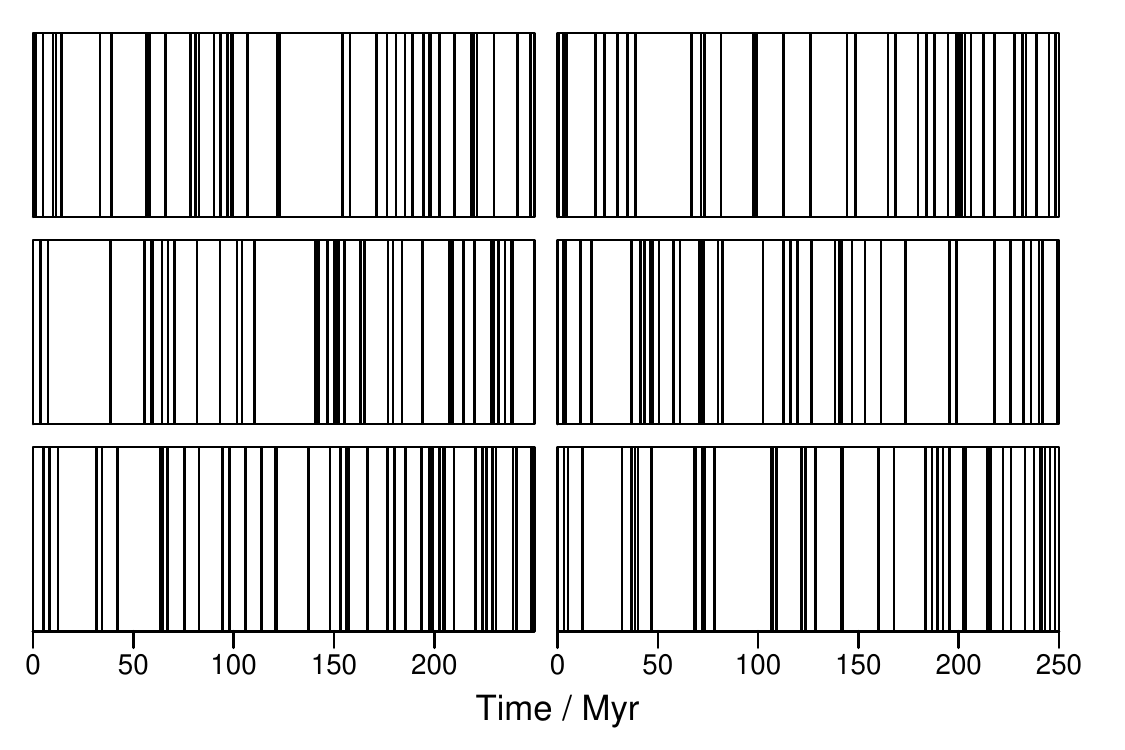} 
\caption{Six examples of simulated time series drawn from the model {\em UniProb} (each with 42 events)
\label{fig:simTS_UniProb}}
\end{center}
\end{figure}

\begin{figure} 
\begin{center}
\includegraphics[width=0.5\textwidth, angle=0]{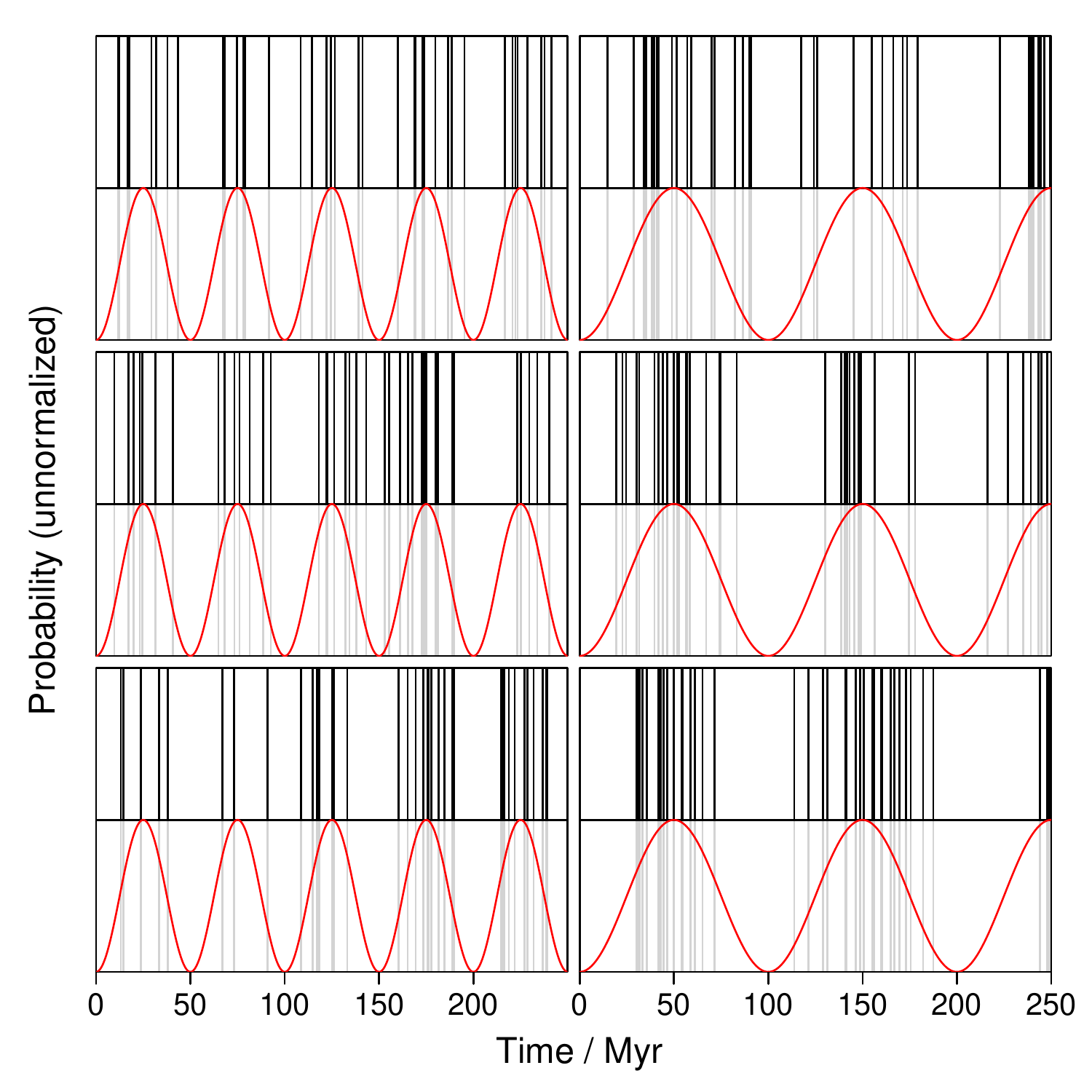} 
\caption{Six examples of simulated time series drawn from the model {\em SinProb} (each with 42 events). The model is plotted in red, the randomly selected events in black (in a split panel to aid viewing of these stochastic time series). The model has a period of 50\,Myr on the left and 100\,Myr on the right (phase\,=\,0.5 in both cases)
\label{fig:simTS_SinProb}}
\end{center}
\end{figure}

\begin{figure} 
\begin{center}
\includegraphics[width=0.5\textwidth, angle=0]{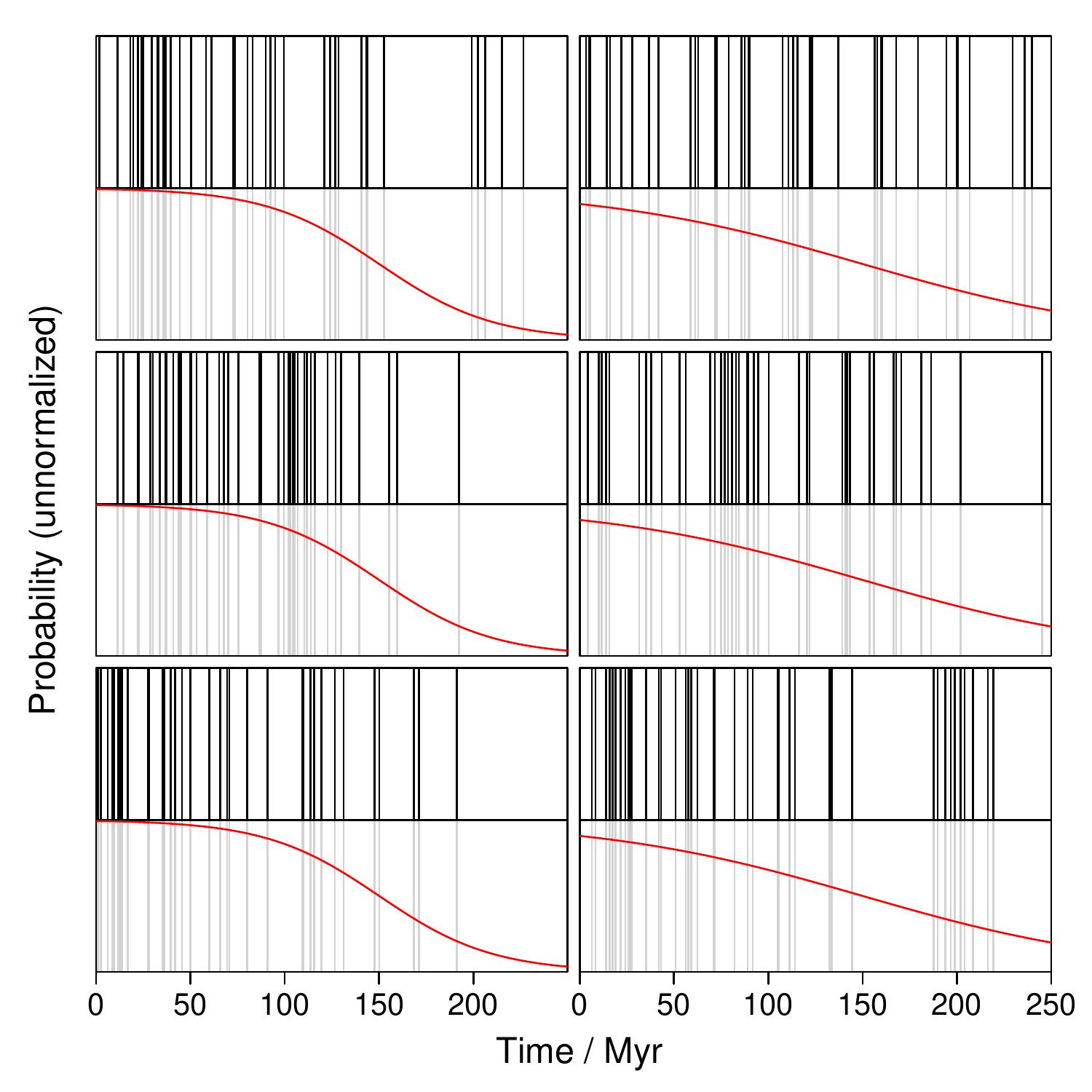} 
\caption{Six examples of simulated time series drawn from model {\em SigProb} (each wth 42 events). The model is plotted in red, the randomly selected events in black (in a split panel to aid viewing of these stochastic time series). The model has $\lambda=-30$ on the left and $\lambda=-70$ on the right ($t_0=150$ in both cases)
\label{fig:simTS_SigProb}}
\end{center}
\end{figure}

Figures~\ref{fig:simTS_UniProb}, \ref{fig:simTS_SinProb} and \ref{fig:simTS_SigProb} show example time series drawn from {\em UniProb}, {\em SinProb} and {\em SigProb} respectively. Note how different the time series can be even when drawn from the same model with the same parameters. It is also interesting -- but not surprising -- that stretches of some of the uniform random distribution appear almost periodic. Conversely, the time series drawn from {\em SinProb} do not always look very periodic. In all cases the distribution can be very uneven and/or show clustering and gaps. Searching for patterns by eye can be quite misleading.

I report here just the results of the periodic models using the uniform prior over period. The results of using a uniform prior over frequency are very similar (I mention a few in the footnotes). Sometimes the evidence is higher or lower, but it does not change the strength of any of the conclusions reported. The maximum likelihood solutions are, in virtually all simulations, identical in parameters and likelihood to within 0.1\%.

\subsection{Periodic data sets}\label{sect:sim_periodic}

I generate data sets from {\em SinProb} at several different fixed periods and phases. 
For each time series I calculate the evidence and Bayes factors (evidence ratios) for a number of models.

\begin{figure} 
\begin{center}
\includegraphics[width=0.5\textwidth, angle=0]{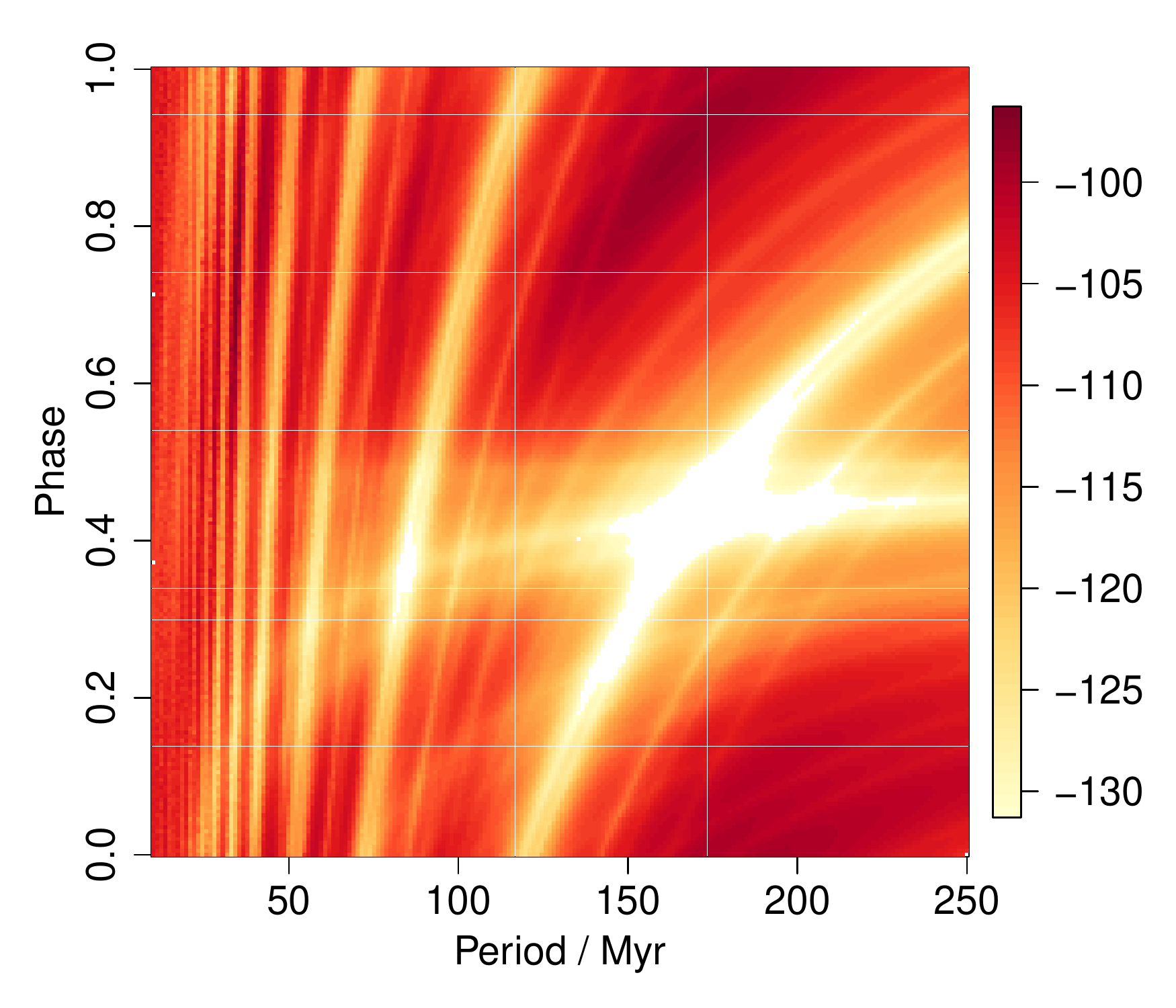} 
\caption{Likelihood distribution as a function of period and phase for the {\em SinProb} model calculated on the simulated data set shown in Fig.~\ref{fig:testrun1003_trueTS} (likelihoods were calculated up 550\,Myr, but are only plotted up to 250\,Myr.). The logarithm (base 10) of the likelihoods are shown on a colour scale: it spans 35 orders of magnitude. White regions are those where the likelihood drops below the minimum plotted.
\label{fig:testrun1003_loglikemap}}
\end{center}
\end{figure}

\begin{figure} 
\begin{center}
\includegraphics[width=0.49\textwidth, angle=0]{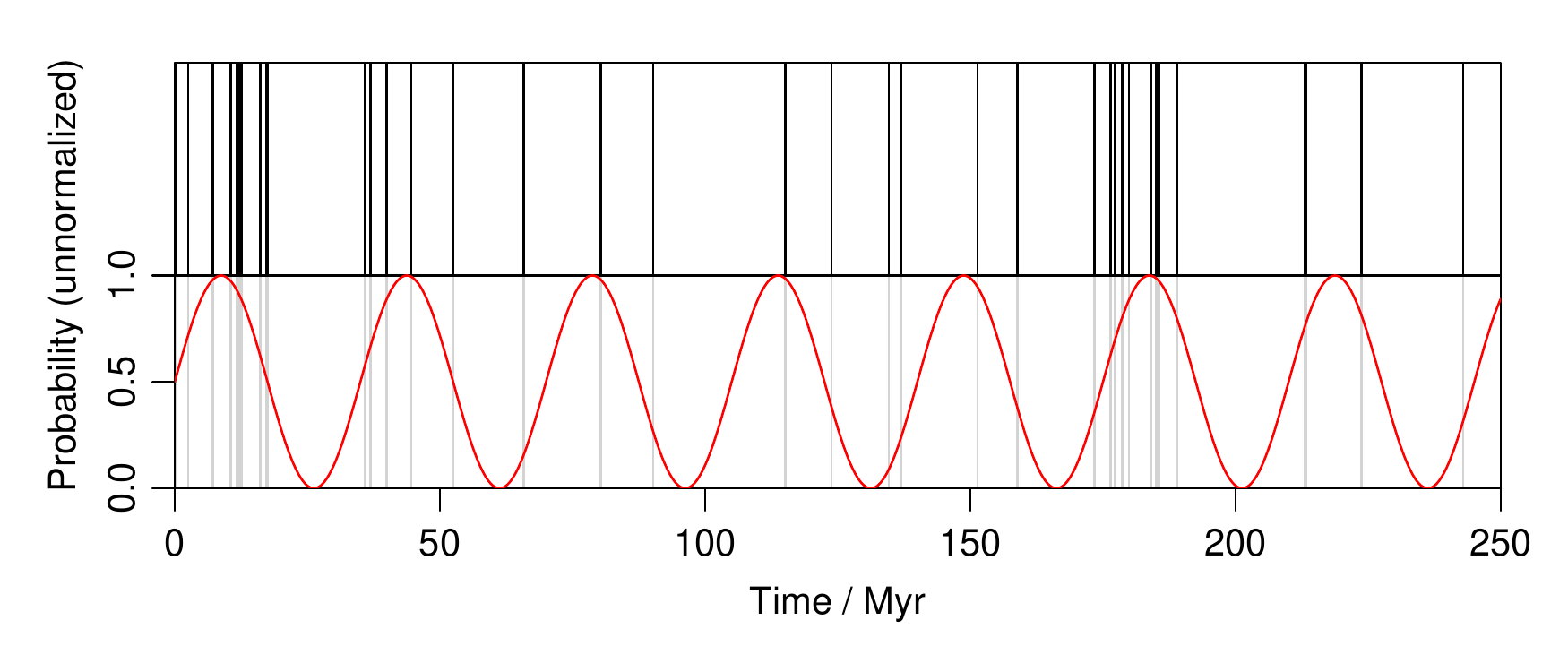} 
\caption{A simulated time series (black lines) drawn from {\em SinProb} with $T=35$\,Myr and $\phi=0.75$ (red curve)
\label{fig:testrun1003_trueTS}}
\end{center}
\end{figure}

For the {\em SinProb} model, the likelihood is calculated at several million values of period and phase drawn from the uniform prior distribution for periods of 10--550\,Myr and phases of 0--1.
These likelihoods are plotted as a density plot in Fig.~\ref{fig:testrun1003_loglikemap} for one particular simulated time series with true period 35\,Myr and phase of 0.75 (shown in Fig.~\ref{fig:testrun1003_trueTS}).  There is large variation in the likelihood. The overall evidence for this model for the full period range 10--550\,Myr is found by averaging the likelihoods, and is $8.12\times10^{-101}$.
The evidence for {\em UniProb} is $8.89\times10^{-102}$, giving a Bayes factor of 9.1. This is not quite significant according to standard criteria, and would appear to suggest a lack of evidence for the periodic model at first. However, we have searched over a wide range of ``periods'': up to twice the time span of the data. We see in Fig.~\ref{fig:testrun1003_loglikemap} that these have very low likelihoods, bringing down the average (the evidence) for {\em SinProb}. So if we instead average over periods of 10--125\,Myr then we get an evidence for this ``properly periodic'' model of $3.22\times10^{-100}$, a Bayes factor of 36 relative to {\em UniProb}. This is good evidence that the periodic model is the better of the two.\footnote{The evidence for {\em SinProb:10:125} with the uniform frequency prior is $3.67\times10^{-100}$, hardly any different.}
This is also true for the model over intermediate periods: BF({\em SinProb10:250}$/${\em UniProb})\,=\,24.5.

\begin{figure} 
\begin{center}
\includegraphics[width=0.40\textwidth, angle=0]{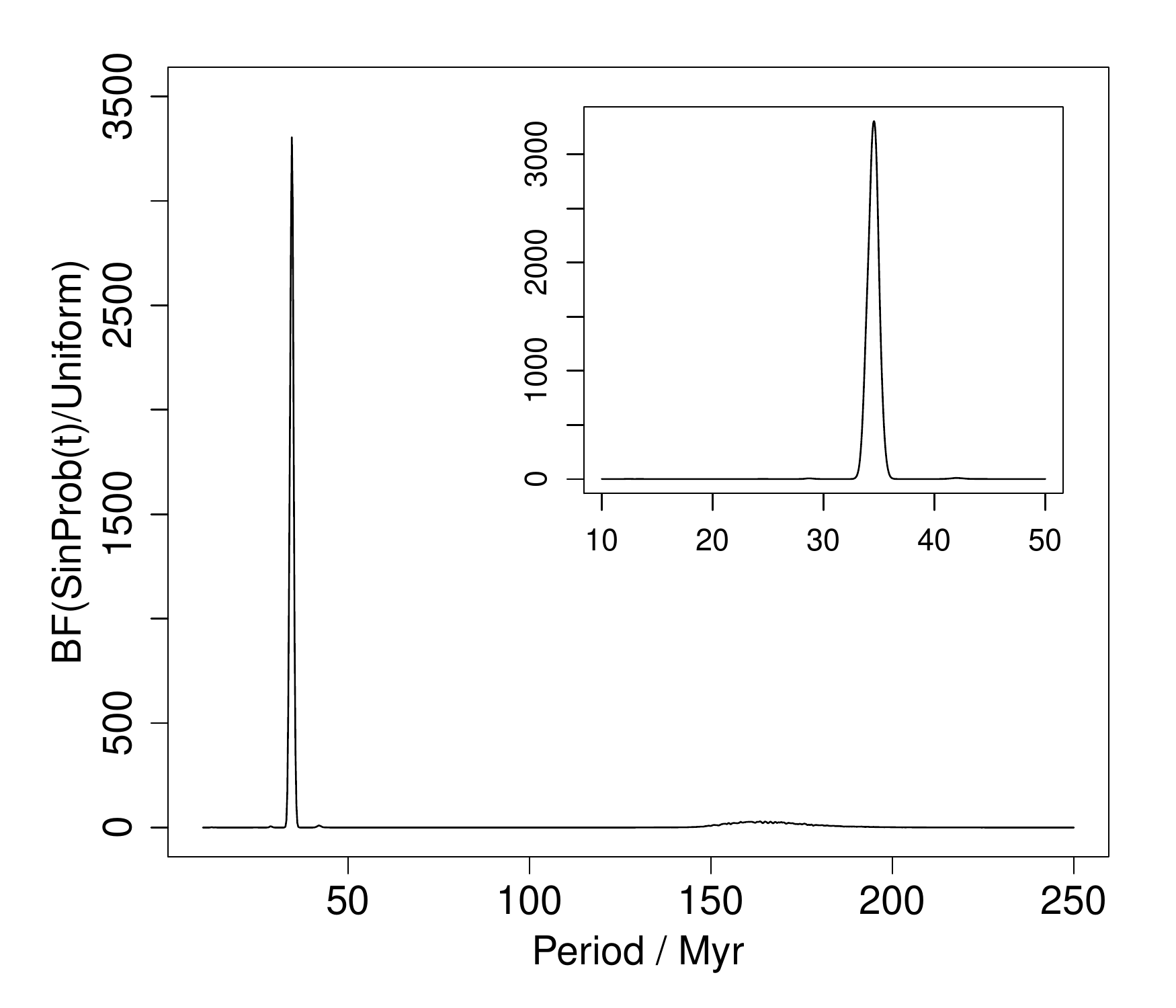}
\caption{Bayesian periodogram obtained by marginalizing the 2D likelihood distribution in Fig.~\ref{fig:testrun1003_loglikemap} over the phase. The inset shows a zoom on the 10--50\,Myr portion
\label{fig:testrun1003_periodogram}}
\end{center}
\end{figure}

We can also calculate the evidence over very narrow ranges of period (for all phases). If we do this for consecutive ranges, we get a {\em Bayesian periodogram}: the variation of likelihood with period (or frequency). This is equivalent to marginalizing the two-dimensional likelihood over phase. It is plotted for the present example in Fig.~\ref{fig:testrun1003_periodogram}. As these likelihoods have uninterpretable absolute values, I divide (``normalize'') the likelihoods by the evidence of the {\em UniProb} model. The periodogram gives the Bayes factor of {\em SinProb} at a specific period (strictly, over a very narrow period range) relative to {\em UniProb}, which I denote as BF({\em SinProb}($t$)$/${\em UniProb}).

We see a (very) significant peak at just a single period, 34.75$\pm 0.6$\,Myr (the uncertainty being the half-width at half maximum, HWHM).  As we have already established evidence for the periodic model overall, this indicates that the periodic model at the detected period is a good explanation for these data.\footnote{The peak in the periodogram for model with the uniform frequency prior is at 34.64\,Myr, and has a BF 86\% as large.}

The maximum likelihood solution is at a period of 34.5\,Myr and a phase of 0.73 
Its likelihood 
is 59\,000 times higher than the evidence for {\em UniProb}.
This is larger than the peak in the periodogram because now we have also found the optimum phase. However, because this model has two parameters which have been fit to the data, this is not a representative measure of the ``significance'' of a periodic model in which the parameters are not known a priori. This will be discussed further below.

This example is actually one in which the simulated time series is relatively non-periodic. Many of the random generations produce much more periodic data sets which achieve peaks in the periodogram of $10^5$ or more and a BF({\em SinProb10:125}$/${\em UniProb}) of hundreds to thousands.

The method is very efficient at finding periods in these data sets.  Of 100 different random time series drawn from {\em SinProb} with period 35\,Myr and phase 0.5, all 100 showed a (very) significant Bayes factor (relative to {\em UniProb}) for both the overall model ({\em SinProb10:125})
and at the peak period (which agreed with true period 
in 99 cases).

I obtain very similar results for numerous other time series simulated at other periods and phases: the evidence for the {\em SinProb10:125} model is almost always very significant, and there is also always a very strong peak in the Bayesian periodogram. 
(As one would expect the peak is wider -- the period less certain -- for longer periods.)
The method can also detect periods between 125 and 250\,Myr, although periods longer than 250\,Myr (where there is less than a complete cycle in the data) are sometimes not detected. Periods down to the lowest limit searched for, 10\,Myr, can also be recovered reliably.


It must be stressed that the above has only established evidence of {\em SinProb} relative to {\em UniProb}. In Bayesian hypothesis testing we we can only ever {\em compare} models. 
I therefore calculated the evidence for the trend model, {\em SigProb}, defined over the parameter ranges
$-100 \leq \lambda \leq +100$, $0 < t_0 < 275$.
For the above example (Fig.~\ref{fig:testrun1003_trueTS}), we get a
Bayes factor BF({\em SinProb10:125}$/${\em SigProb})\,=\,155, clearly favouring the periodic model.
This is also seen for other simulated time series at the same and other periods: the Bayes factor BF({\em SinProb}$/${\em SigProb}) is
typically 10--1000 for input periods below 250\,Myr. For longer periods the {\em SigProb} model sometimes dominates, depending on the exact data set. This is because such long ``periods'' are really trends, and these may be fit better with {\em SigProb}.

In conclusion, if the time series is periodic (in the sense of a periodic event probability), then the method
strongly favours the periodic model over both a uniform one and a monotonic trend model. We conclude this
primarily from the significant (large) Bayes factor for the model as a whole (i.e.\ a broad period range for all phases), and then from a significant peak in the Bayesian periodogram (the Bayes factor over a very narrow period range).



\subsection{Uniform and trend data sets}\label{sect:sim_trend}

Let us now examine whether we erroneously favour the {\em SinProb} model -- and possibly detect artificial periods --
in non-periodic data. 

\subsubsection*{Uniform data sets}
For this purpose I simulated 50 stochastic time series from the {\em UniProb} model and calculated the evidence
for both {\em UniProb} and {\em SinProb}. In none of these 50 cases is BF({\em SinProb10:125}$/${\em UniProb}) significant: most of the time series
give $10^{-2}$ to $10^{-3}$ (the full range is $3\times10^{-4}$ to 1.1).
BF({\em SinProb10:250}$/${\em UniProb}) is likewise
much less than one.
We would therefore correctly conclude that a uniform random distribution is a much
better model than a periodic one. (If no other model were plausible, we could further conclude with some confidence that this is the ``best'' model.)

However, in 10 of the cases we nonetheless observe a significant peak in the periodogram at periods less than 125\,Myr, where significant means BF({\em SinProb}($t$)$/${\em UniProb})\,$>$\,10. This is because even a uniform random time series with 42 points can happen to show a weak periodicity at {\em some} period, even though the periodic model overall is a poorer explanation for the data than the uniform one.\footnote{Half of the time series in fact show a peak in the periodogram with BF({\em SinProb}($t$)$/${\em UniProb}) in the range 1 to 10. We clearly must resist the temptation to read too much into such low significance peaks!}  If we took just the identification of these ``significant'' peaks to be evidence for a period (as is done in frequentist periodogram analyses), we would make a false positive claim in 20\% of cases.  This underlines the importance of basing a conclusion on the evidence for the model as a whole, rather than a specific fit. 

\subsubsection*{Trend data sets}
We see the same general phenomena among a set of time series generated from {\em SigProb}: The evidence for {\em SinProb10:125} is much lower than the evidence for {\em SigProb}, leading us to correctly conclude a lack of evidence for periodicity. But once again there are spurious ``significant'' peaks in the periodogram.
The reason for this is that although {\em SinProb} sometimes gives high evidence at a very specific period, once averaged over a broader period the evidence is much less. If the peak had been much higher (as was the case for the truly periodic data sets above, e.g.\ Fig.~\ref{fig:testrun1003_periodogram}), then averaging over a broader period reduces the evidence, but not enough to make it insignificant.  The model evidence is a balance between the quality of a specific fit, and how much of the parameter space produces a good fit. (This is discussed further in section~\ref{sect:discussion}, where it can be understood in terms of Occam factors).  As we had no prior reason to suspect a periodicity {\em at the peak found}, it is misguided to focus only on that peak -- and that may anyway give the wrong conclusion, as we have just seen.

\begin{figure} 
\begin{center}
\includegraphics[width=0.49\textwidth, angle=0]{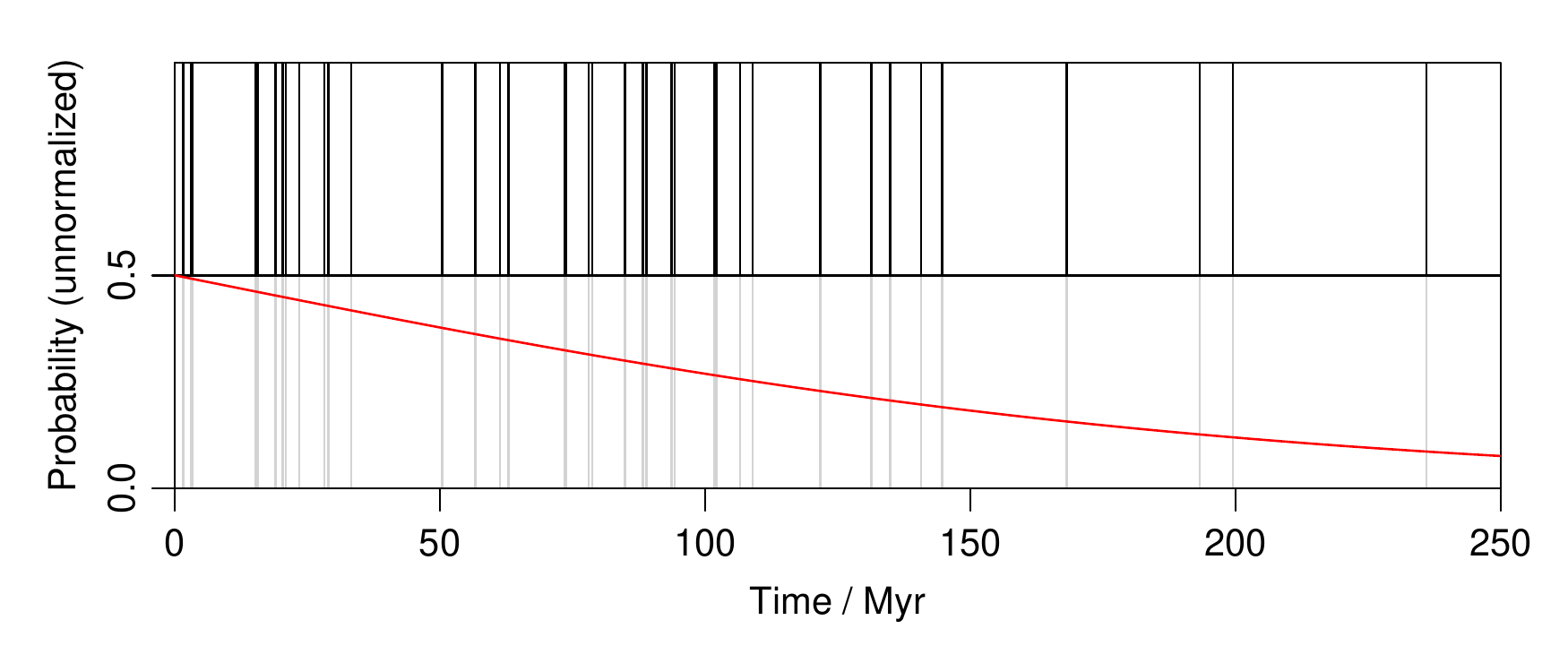} 
\caption{A simulated time series (black lines) drawn from {\em SigProb} with $\lambda=-100$\,Myr and $t_0=0$\,Myr (red curve)
\label{fig:testrun1022_trueTS}}
\end{center}
\end{figure}

\begin{figure} 
\begin{center}
\includegraphics[width=0.5\textwidth, angle=0]{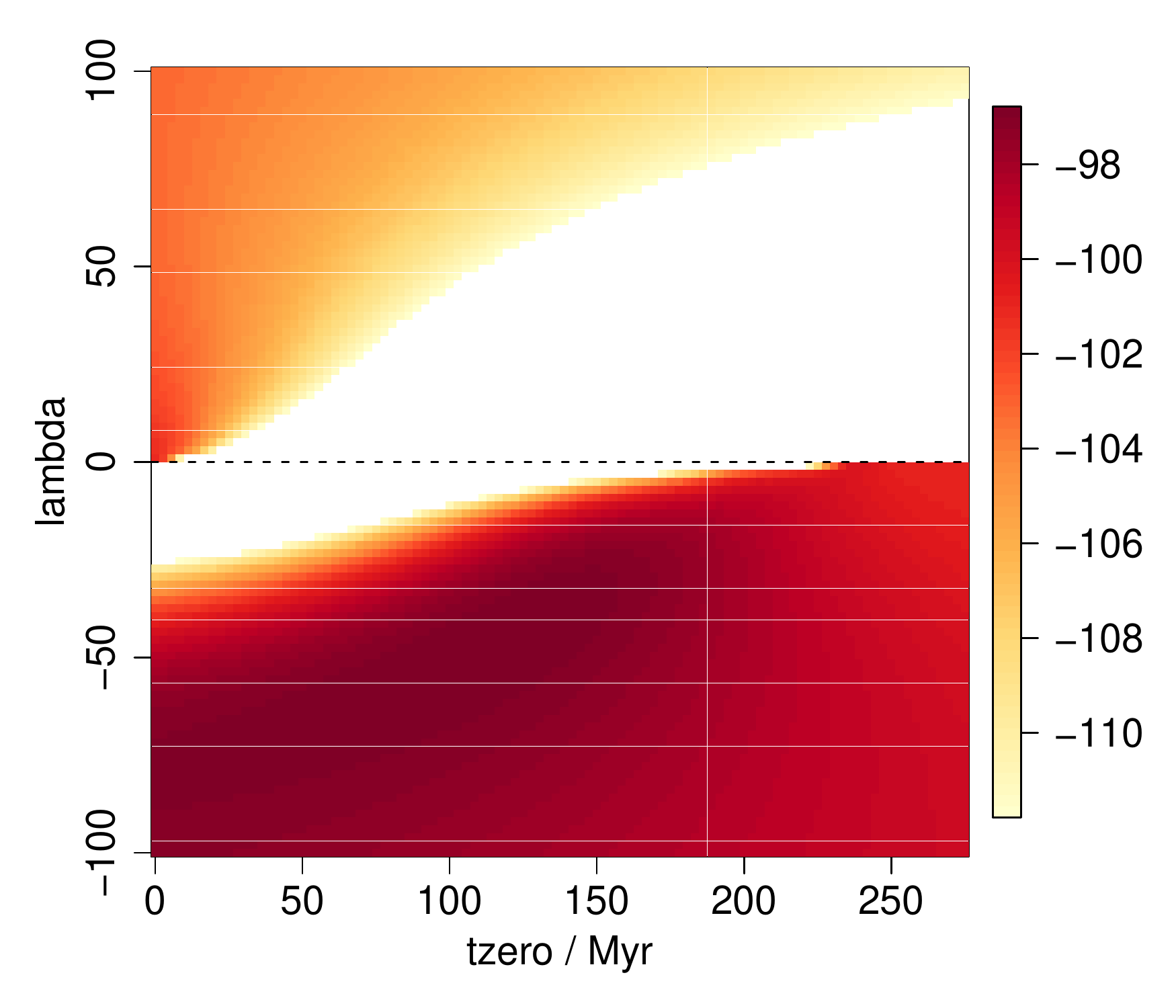} 
\caption{Log likelihood distribution as a function of $\lambda$ and $t_0$ in the {\em SigProb} model calculated for the simulated data set shown in Fig.~\ref{fig:testrun1022_trueTS}. 
The log(Evidence) for the overall model is $-97.8$ (and 0.3 higher for {\em SigProb:Neg}).
For comparison, log(Evidence)\,=\,$-100.9$ for {\em UniProb}.
\label{fig:testrun1022_loglikemap}}
\end{center}
\end{figure}

An example time series drawn from the trend model is shown in Fig.~\ref{fig:testrun1022_trueTS}.
The likelihood distribution for the {\em SigProb} model
is shown in Fig.~\ref{fig:testrun1022_loglikemap}, with a colour scale spanning 15 orders of magnitude. Positive trends 
(probability increasing with look back time) are heavily disfavoured. In contrast, a very broad range of parameter space for $\lambda<0$ has large evidence: the model is not very sensitive to the exact parameters. This contrasts with the simulated periodic data sets, where the evidence for the overall model came from a very narrow range of the parameters 
with very high likelihood (see Figs~\ref{fig:testrun1003_loglikemap} and~\ref{fig:testrun1003_periodogram}).

It is interesting to note that 
the evidence for {\em SinProb10:125} on these trend data sets is low even when compared to {\em UniProb}: The
Bayes factors range from 0.7 to $5\times10^{-5}$ in 39 of 40 simulations (the other value was 2).
So even when comparing against an overly simple model, we get no evidence for periodicity. 
Yet the in the {\em periodograms} normalized to the evidence in {\em UniProb} we see apparently significant peaks, i.e.\ with BF({\em SinProb}($t$)$/${\em UniProb})\,$>$\,10.
These peaks are irrelevant, however, not only because {\em SinProb} as a whole is disfavoured but also because {\em UniProb} is a poor model for the data. That is, we can increase the {\em apparent} significance of a model (here {\em SinProb}) by comparing it with an overly-simple alternative model (here {\em UniProb}), a common mistake in hypothesis testing.

\subsection{Compound data sets}\label{sect:compound}

It is plausible that the impact cratering phenomenon comprises both a periodic and a non-periodic component (e.g.\ Grieve et al.\ 1988, Lyytinen et al.\ 2009). It is generally more difficult to identify evidence for more complex models given a limited amount of data and the stochastic nature of the phenomenon.

This difficulty is confirmed by simulations. I simulated 48 time series drawn from {\em SinBkgProb} for several periods between 10 and 250\,Myr and $b=1$, i.e.\ equal amplitudes of the periodic and uniform components (Table~\ref{tab:tsmodels}). In many cases the evidences for {\em SinBkgProb10:550}, {\em SinBkgProb10:125} and {\em UniProb} are about equal, so the true model is not favoured over a pure ``background'' model. 
({\em SigProb} achieves a very low evidence in comparison: we do not erroneously claim a trend when there is not one.)

We see much the same for several time series drawn from {\em SinSigProb} at each of three different periods (20, 35, 50\,Myr) with a fixed trend ($\lambda=-60, t_0=100$). The Bayes factors of the true models relative to just the trend models, {\em SigProb}, lie between 0.5 and 7.\footnote{When using a uniform prior over frequency the Bayes factors come out to within about 20\% of the same values.} Thus although the true model is not explicitly disfavoured, these values are not high enough to claim significant evidence to favour it, and when lacking evidence in favour of a more complex model, we would probably prefer the simpler one.


It seems that the method is conservative, having difficulties to identify a period in a stochastic time series which includes a large (here 50\% amplitude) non-periodic component.  A similar conclusion was reached by Lyytinen et al.\ (2009) using a different approach.  However, I have only performed a few tests with these compond data sets, and only for a very narrow part of the input parameter space.  As the sensitivity is likely to vary over the input parameter space, more extensive tests are necessary.

\vspace*{1em}
\noindent
In conclusion of this section on simulations, I have found that if the time series is a stochastic one drawn from a model with a periodic distribution, then there is both significant evidence for periodicity and we can identify the true period. Conversely, if data are drawn from a uniform or trend distribution, we find significant evidence for the correct model, and do not erroneously identify periodicity. Preliminary tests indicate difficulty in favouring more complex compound models (when true) over a simpler one. It may be that 42 events is simply inadequate to support more complex models.


\section{Results}\label{sect:results}

\begin{table*}
\begin{center}
\caption{The evidence (equation~\ref{eqn:evidence}) for different models and data sets. The Bayes factor is the ratio of two evidences (for a given data set). See the text for the exact parameter ranges used in each case.
\label{ressum}}
\vspace*{1em}
\begin{tabular}{l e{4} e{4} e{4} e{4} e{4} e{4} }
\toprule
             & \multicolumn{1}{c}{\rm basic150} & \multicolumn{1}{c}{\rm full150} & \multicolumn{1}{c}{\rm basic250} & \multicolumn{1}{c}{\rm ext250} & \multicolumn{1}{c}{\rm full250} & \multicolumn{1}{c}{\rm large400} \\
\midrule
UniProb               & 1.63e-70 & 2.30e-105 & 1.03e-103 & 1.68e-113 & 3.27e-145  & 1.25e-47 \\
\midrule
SinProb10:50        & 2.67e-71 & 1.41e-105 & 2.87e-104 & 1.11e-113 & 4.83e-145  & 8.89e-48 \\
SinProb10:125      & 1.07e-71 &  5.56e-106 & 1.15e-104 & 4.18e-114 & 1.71e-145  & 1.01e-47  \\
SinProb10:300      & 9.64e-72 &                &                 &                &                 &                 \\
SinProb10:400      &              &                &                 &                &                 & 3.35e-48   \\
SinProb10:550      &               &                & 8.57e-103 &                &                 &                  \\
\midrule
SinBkgProb10:50   & 1.62e-70 &                & 1.11e-103 &  2.23e-113 & 5.22e-145  &                  \\
SinBkgProb10:125 & 1.58e-70 &                & 9.71e-104 &  1.79e-113 & 3.91e-145 &                  \\
SinBkgProb10:300 & 1.74e-70 &                &                 &                &                 &                  \\
\midrule
SigProb                & 1.35e-70 & 1.28e-103 & 8.64e-102  & 1.21e-110 & 7.66e-139 & 2.93e-48      \\
SigProb:Neg          & 2.64e-70 & 2.56e-103 & 1.73e-101 &  2.41e-110 & 1.53e-138 & 4.83e-48      \\
SigProb:Pos           & 5.70e-72 & 3.73e-108 & 1.47e-106 & 8.72e-117 &  3.01e-149 & 1.03e-48     \\
\midrule
SinSigProb            &               &                &  6.94e-102 & 6.81e-111 & 7.97e-140 &                 \\
\bottomrule
\end{tabular}
\end{center}
\end{table*}

Armed with the experience of how the models respond to time series of known origin, I now turn to analysing the real cratering data.  
For quick refefence, the results are summarized in Table~\ref{ressum}. For the periodic models I only report results using a prior uniform over period, because the results for using a  prior uniform over frequency are the same in all relevant respects. An example is shown in the appendix.

I should point out that although I calculate likelihoods for the periods models for periods up to twice the time span, I do this only to illustrate that long ``periods'' can only really be interpreted as long-term trends, and so are better modelled by the {\em SigProb} model.
The main results for ``periodicity'' are for the evidence calculated over narrower period ranges.

\subsection{Data sets: basic150 and ext150}

\begin{figure} 
\begin{center}
\includegraphics[width=0.5\textwidth, angle=0]{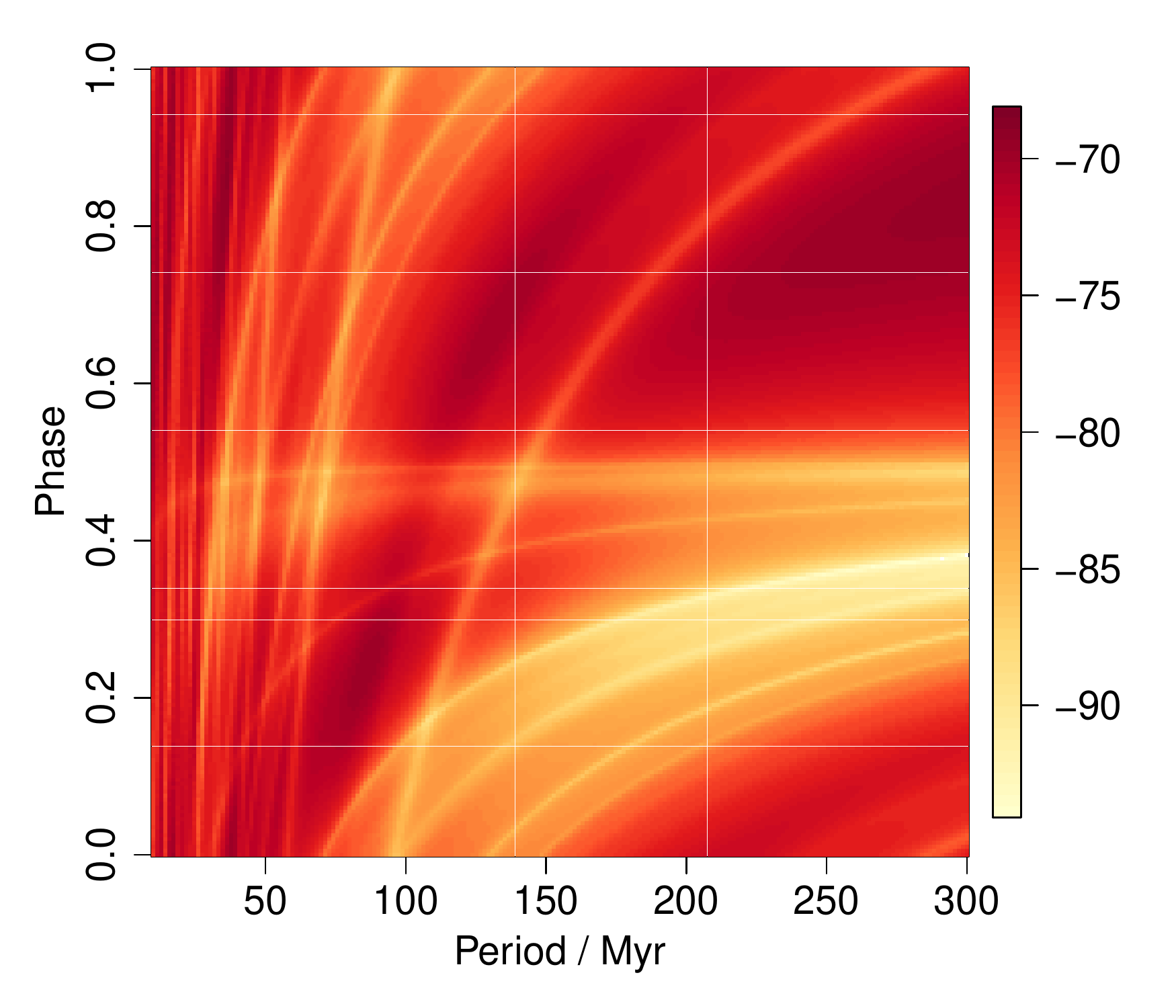}
\caption{Log likelihood distribution as a function of period and phase for the {\em SinProb} model for the basic150 data set
\label{fig:eidrun42_loglikemap}}
\end{center}
\end{figure}

I start with the basic150 data set, the 32 craters over the past 150\,Myr.  The likelihood distribution for the {\em SinProb} model is shown in Fig.~\ref{fig:eidrun42_loglikemap}. The evidence for {\em SinProb10:300} is $9.64\times10^{-72}$. This compares to $1.63\times10^{-70}$ for the {\em UniProb} model, giving a Bayes factor of 0.060, or significant evidence in favour of {\em UniProb}.  The maximum likelihood over this parameter range occurs at (period, phase)\,=\,$(11.7, 0.73)$ with a likelihood of $8.00\times10^{-69}$, or BF({\em SinProb}($t$)$/${\em UniProb})\,=\,49.  Although this tells us that this very specific fit explains the data better than a uniform random distribution, it only comes about because we have tuned two parameters (period and phase). {\em UniProb} has no free parameters and cannot be tuned, so a simple comparison of maximum likelihood does not take into account the model complexity. This is why we must use the evidence for the models as a whole. After all, we wanted to look for evidence of periodicity in general, not for evidence of ``a period of 11.7\,Myr and phase 0.73'', and more complex models could be defined which for which there are even higher likelihood fits.

Limiting the evidence calculation to shorter periods of 10--50\,Myr, we get a Bayes factor of {\em SinProb10:50} relative to {\em UniProb} of 0.16, still insignificant evidence for periodicities.

\begin{figure} 
\begin{center}
\includegraphics[width=0.40\textwidth, angle=0]{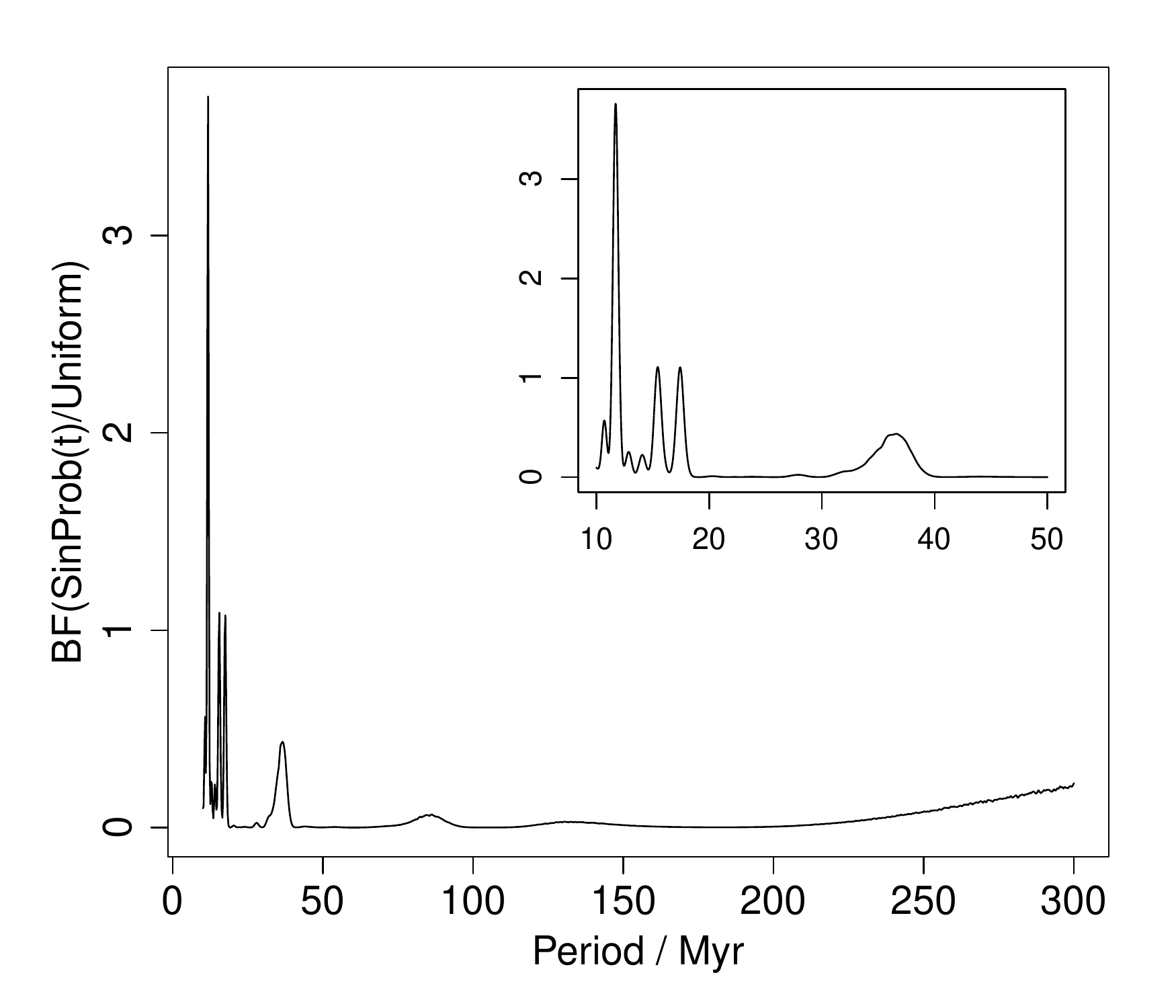}
\caption{Bayesian periodogram for the {\em SinProb} model for the basic150 data set
\label{fig:eidrun42_periodogram}}
\end{center}
\end{figure}

The Bayesian periodogram (section~\ref{sect:sim_periodic}) is shown in Fig.~\ref{fig:eidrun42_periodogram}. There is no significant evidence for periodicity at any period.  Recall that, in the simulations, we obtained a peak (at the true period) with far higher Bayes factors (significance) when the data were drawn from a sinusoidal probability distribution.

In the appendix I repeat this analysis using a prior uniform in frequency for {\em SinProb}.

If the impact history comprises both constant and periodic components, then a model which reflects this may identify a periodicity better.
To examine this I calculate likelihoods for {\em SinBkgProb}, which adds a variable constant background term, $b$, to {\em SinProb}. For the full
period range and for $0 \leq b \leq 5$ the evidence is $1.74\times10^{-70}$, a Bayes factor relative to {\em UniProb} of 1.1. 
If we limit the period range to 10--50\,Myr this rises only to 1.2. The periodogram shows smaller Bayes factors than in Fig.~\ref{fig:eidrun42_periodogram}, even when calculated for limited ranges of $b$.
Hence {\em SinBkgProb} describes the data no better than a purely uniform random distribution, which we should arguably then prefer (see section~\ref{sect:discussion}).


So far the uniform random distribution describes the data as well as or better than a periodic one. This does not mean that this is {\em the} best model, however: we can only assess what we explicitly test. To look for the evidence for a trend in the data, I calculate the evidence for the {\em SigProb} over the parameter range $-100 \leq \lambda \leq +100$, $0 < t_0 < 150$. 
This gives $1.35\times10^{-70}$, a Bayes factor relative to {\em UniProb} of 0.83. 
Splitting {\em SigProb} into two distinct models, one for $\lambda<0$ ({\em SigProb:Neg}) and the other for $\lambda>0$ ({\em SigProb:Pos}), the Bayes factors are 1.63 and 0.035 respectively. The latter -- an increase in impact probability with look back time -- is disfavoured. 

Performing the same analyses on the ext150 data set gives very similar results: Adding these four events does not make the evidence for periodicity nor for a trend significant. 

In summary, we have no evidence for periodicity in the basic150 or ext150 data sets. Of the models tested, both {\em UniProb} and {\em SigProb:Neg} are more or less equally probable explanations. Given a lack of strong evidence in favour of the more complex trend model, I conclude that the simpler, 
uniform random distribution is an adequate -- and plausible -- explanation for impact craters over the past 150\,Myr.

\subsection{Data set: full150}

I now add the 12 craters which have upper ages limits below 150\,Myr, using the approach explained in section~\ref{sect:censored}.
The evidence for the models tested (same parameter ranges for basic150) are listed in Table~\ref{ressum}.
There is now significantly more evidence for the trend model than for the uniform one. More specifically, the negative trend ({\em SigProb:Neg}) is hugely favoured over the positive one.\footnote{As the evidence for {\em SigProb} is the average of the evidence for its two components, and one is five orders of magnitude smaller than the other, then the evidence for {\em SigProb:Neg} is just twice that of {\em SigProb}.}  As the new data are {\em upper} age limits, it is not surprising that their inclusion increases the evidence for {\em SigProb:Neg}, although the evidence is very strong: BF({\em SigProb:Neg}$/${\em UniProb})\,=\,111.

Thus adding the 12 craters with upper age limits in a conservative manner (a flat probability distribution) has made {\em SigProb:Neg} much more probable than {\em UniProb}. {\em SinProb} remains unlikely.

Let us now extend the data set back to 250\,Myr BP.

\subsection{Data set: basic250}

\begin{figure*} 
\begin{center}
\includegraphics[width=0.9\textwidth, angle=0]{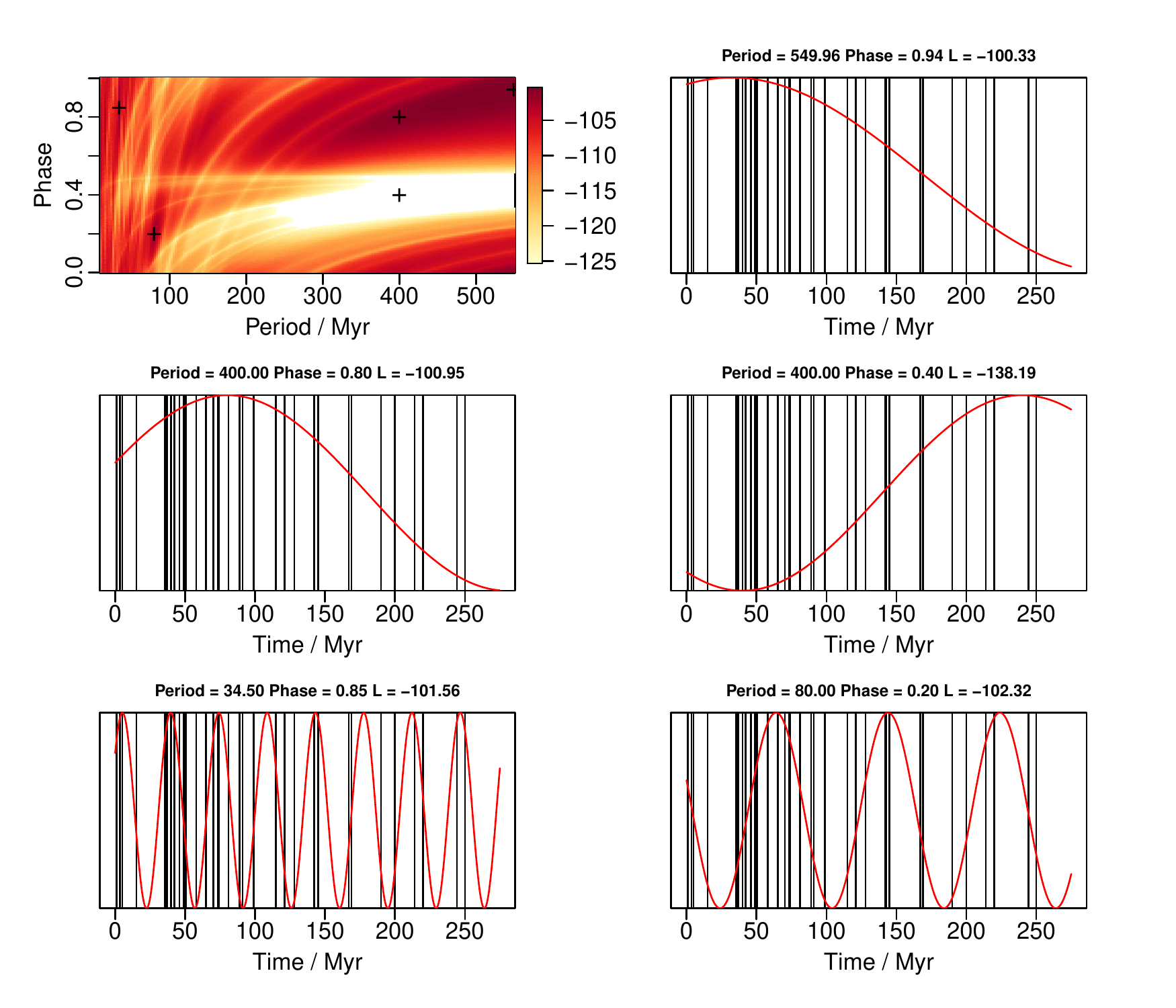}
\caption{The top-left panel shows the log likelihood distribution as a function of period and phase for the {\em SinProb} model for the basic250 data set. The red curves in the five other panels show five possible ``solutions'' at five $(T,\phi$) values corresponding to the five crosses plotted on the likelihood distribution. The black lines are the basic250 data. The log (base 10) likelihood and parameters of each solution are given at the top of each panel. The top-right panel is the maximum likelihood solution. The bottom left-panel is (close to) the maximum likelihood solution for short periods. For comparison, log(Evidence) for {\em UniProb} and {\em SigProb:Neg} are $-102.99$ and $-100.76$ respectively.
\label{fig:eidrun43_SinProbmodels}}
\end{center}
\end{figure*}

\begin{figure} 
\begin{center}
\includegraphics[width=0.40\textwidth, angle=0]{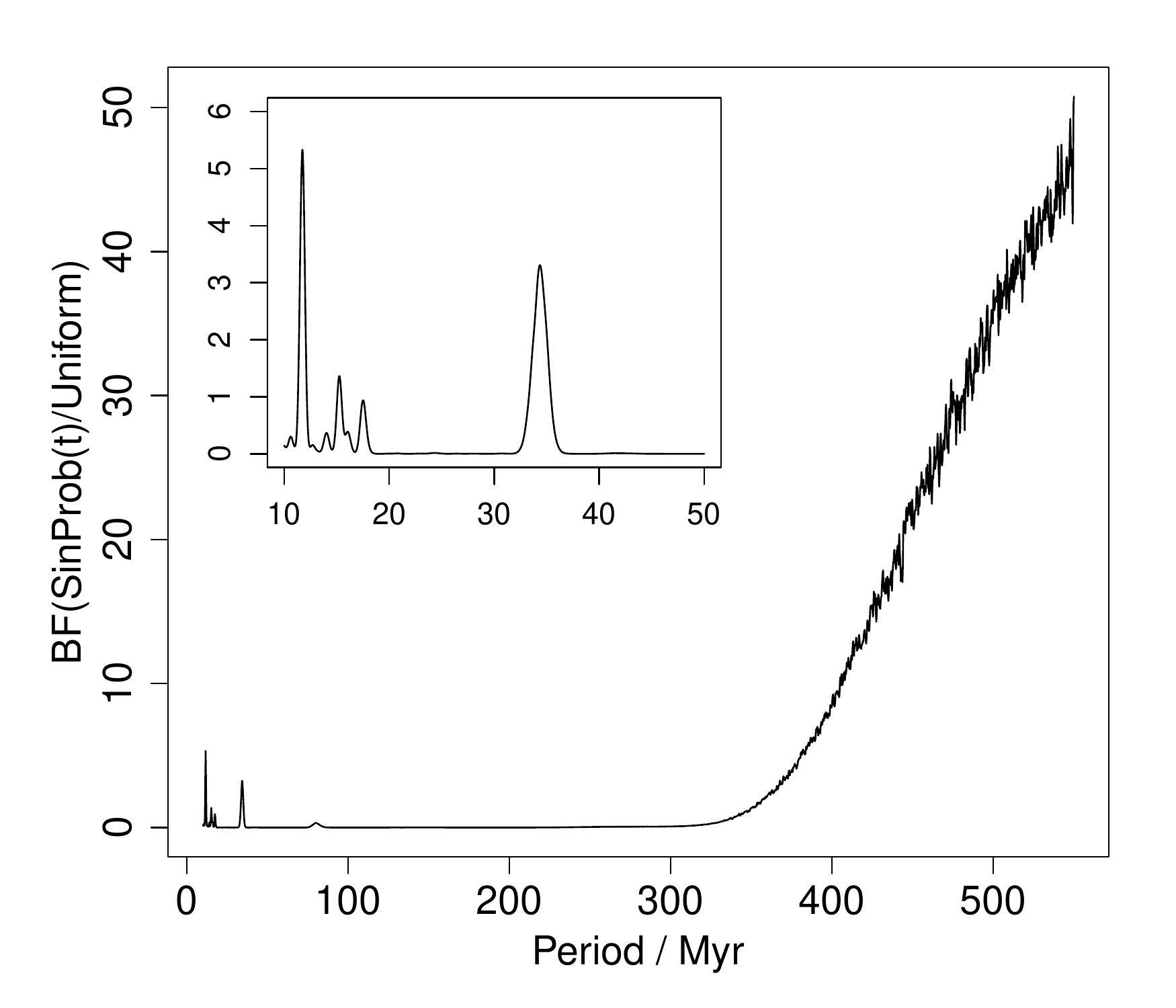}
\caption{Bayesian periodogram for the {\em SinProb} model for the basic250 data set. If we normalize the periodogram against model {\em SigProb:Neg} instead of the {\em UniProb} then the Bayes factors are reduced by a factor of 167.
\label{fig:eidrun43_periodogram}}
\end{center}
\end{figure}

This data set comprises 42 events.  The evidence for {\em UniProb} is $1.03\times10^{-103}$ compared to $8.57\times10^{-103}$ for {\em SinProb10:550}, a Bayes factor of 8.3.  The top-left panel of Fig.~\ref{fig:eidrun43_SinProbmodels} shows the likelihood distribution. The largest likelihoods are in the upper right quadrant, for periods greater than 200\,Myr and phases above 0.5. Indeed, the maximum likelihood solution is at a period of 550\,Myr with a phase of 0.94: it is plotted in the top-right panel of the same figure. This ``period'' is twice the duration of the data and is actually modelling a trend of decreasing probability with look back time. Marginalizing the likelihood distribution over all phases to produce the periodogram (Fig.~\ref{fig:eidrun43_periodogram}), we see that all of the significant periods are at these very long trend periods.  The evidence for true periods -- {\em SinProb10:125} -- is $1.15\times10^{-104}$, a Bayes factor relative to {\em UniProb} of 0.11.

Given that the likelihood varies considerably across the parameter space, it is informative to examine the model at various parameter ``solutions''. Five examples are shown in Fig.~\ref{fig:eidrun43_SinProbmodels}. The two panels in the central row have the same period but different phases: the right-hand one gives an {\em increasing} probability of events with look back time and is hugely disfavoured by the data compared to the decreasing trend.
The lower two panels are local maxima in the likelihood distribution at shorter periods. They look as though they could reasonably produce the observed data, bearing in mind that these are stochastic models. However, even the better of the two has a likelihood 10--100 times smaller than the longer period ``trend'' solutions.

Introducing a constant background into the model ({\em SinBkgProb}), we again find that the highest likelihood solutions (and the only ones more likely than {\em UniProb}) are at long ``periods''.  Solutions with $b>0.5$ are favoured compared to those with a smaller (or zero) background.  There is no evidence for any proper period ($<125$\,Myr), with Bayes factors relative to {\em UniProb} of no more than 0.94, even if we examine narrow ranges of $b$. 
So there is no evidence for periodicity superimposed on a constant background probability (although recall from section~\ref{sect:compound} that it may be hard to identify such a model with confidence).

\begin{figure} 
\begin{center}
\includegraphics[width=0.5\textwidth, angle=0]{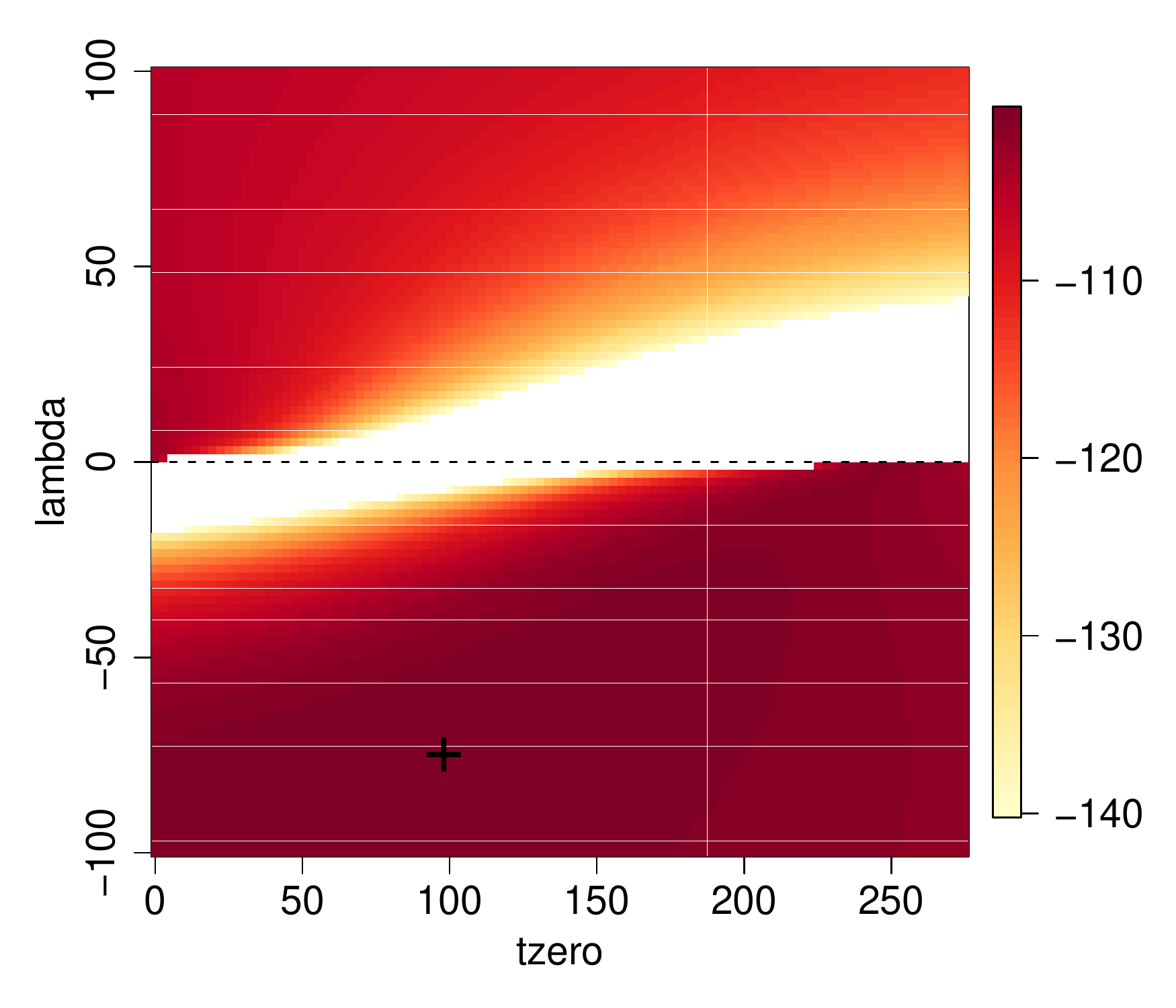}
\caption{Log likelihood distribution as a function of $\lambda$ and $t_0$ for the {\em SigProb} model for the basic250 data set. The black cross marks the maximum likelihood solution.
\label{fig:eidrun56_loglikemap}}
\end{center}
\end{figure}

Given the above evidence for a trend, I fit {\em SigProb} over the parameter range $-100 \leq \lambda \leq +100$, $0 < t_0 < 275$. 
The overall evidence is $8.64\times10^{-102}$. This includes both positive and negative trends:
Fig.~\ref{fig:eidrun56_loglikemap} shows that the former ($\lambda>0$) have much lower likelihoods. 
Splitting this model into two, we find that the evidence for $\lambda<0$ ({\em SigProb:Neg}) is $1.73\times10^{-101}$, a Bayes factor of 167 relative to {\em UniProb}. 

As {\em SigProb:Neg} is much more plausible than {\em UniProb}, this tells us that {\em UniProb} is an inappropriate reference model for the periodogram in Fig.~\ref{fig:eidrun43_periodogram}.  By comparing the evidence at each period to a model which predicts the data poorly, the periodic model may of course look good in comparison.  But this is not the same as saying that the period is significant, because other (plausible) models may predict the data even better, as is the case here. This mistake is often made when assessing the significance of the classical (e.g.\ Lomb--Scargle) periodogram.  One must be careful to choose an appropriate ``background'' or ``reference'' model for comparison.  When using the evidence for {\em SigProb:Neg} as the reference in the periodogram, the Bayes factors in Fig.~\ref{fig:eidrun43_periodogram} are all reduced by a factor of 167, rendering all peaks -- and even the long ``trend'' periods -- entirely insignificant. Overall, the negative trend model is far more probable than the periodic one: BF({\em SigProb:Neg}$/${\em SinProb10:125})\,=\,1500.

\begin{figure} 
\begin{center}
\includegraphics[width=0.5\textwidth, angle=0]{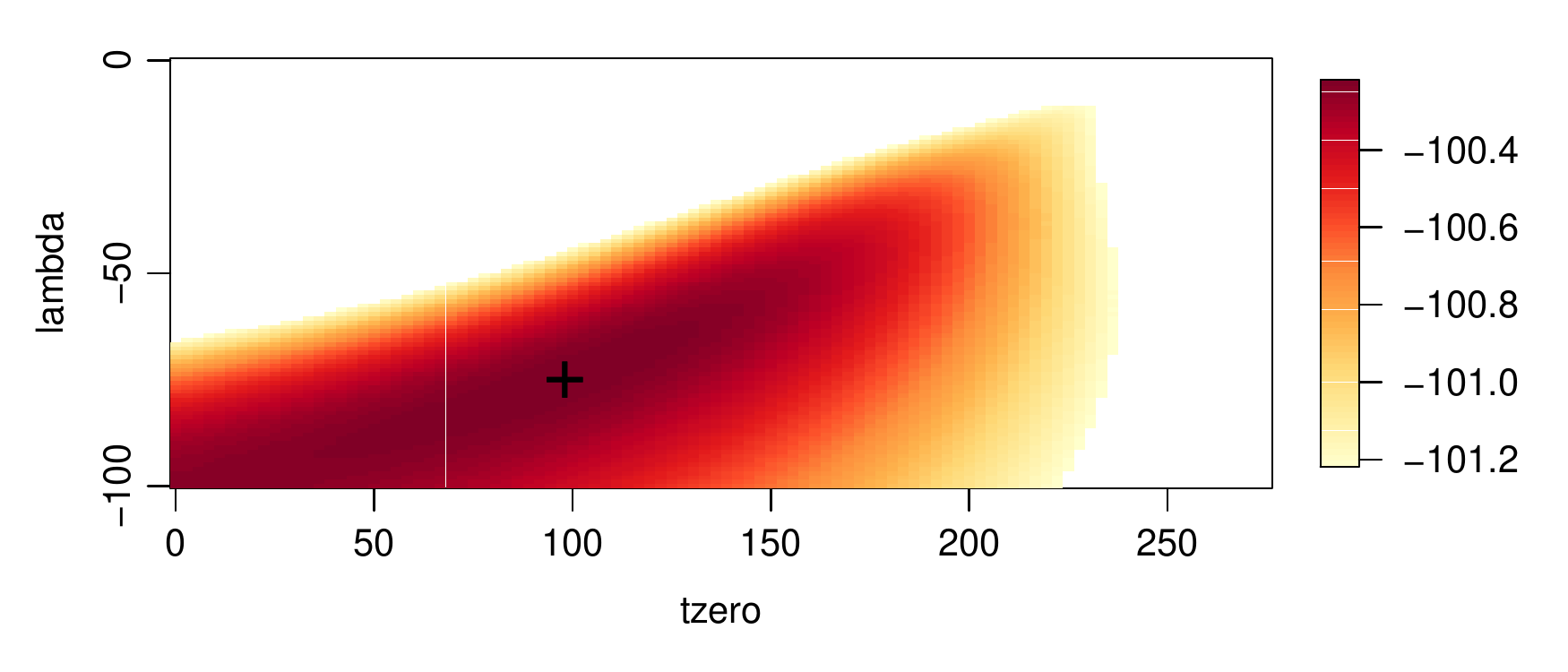}
\caption{As Fig.~\ref{fig:eidrun56_loglikemap} but only for $\lambda<0$ and with a shallower likelihood scale to better show the region around the peak. 
\label{fig:eidrun56_loglikemap_MLpeak}}
\end{center}
\end{figure}

\begin{figure} 
\begin{center}
\includegraphics[width=0.4\textwidth, angle=0]{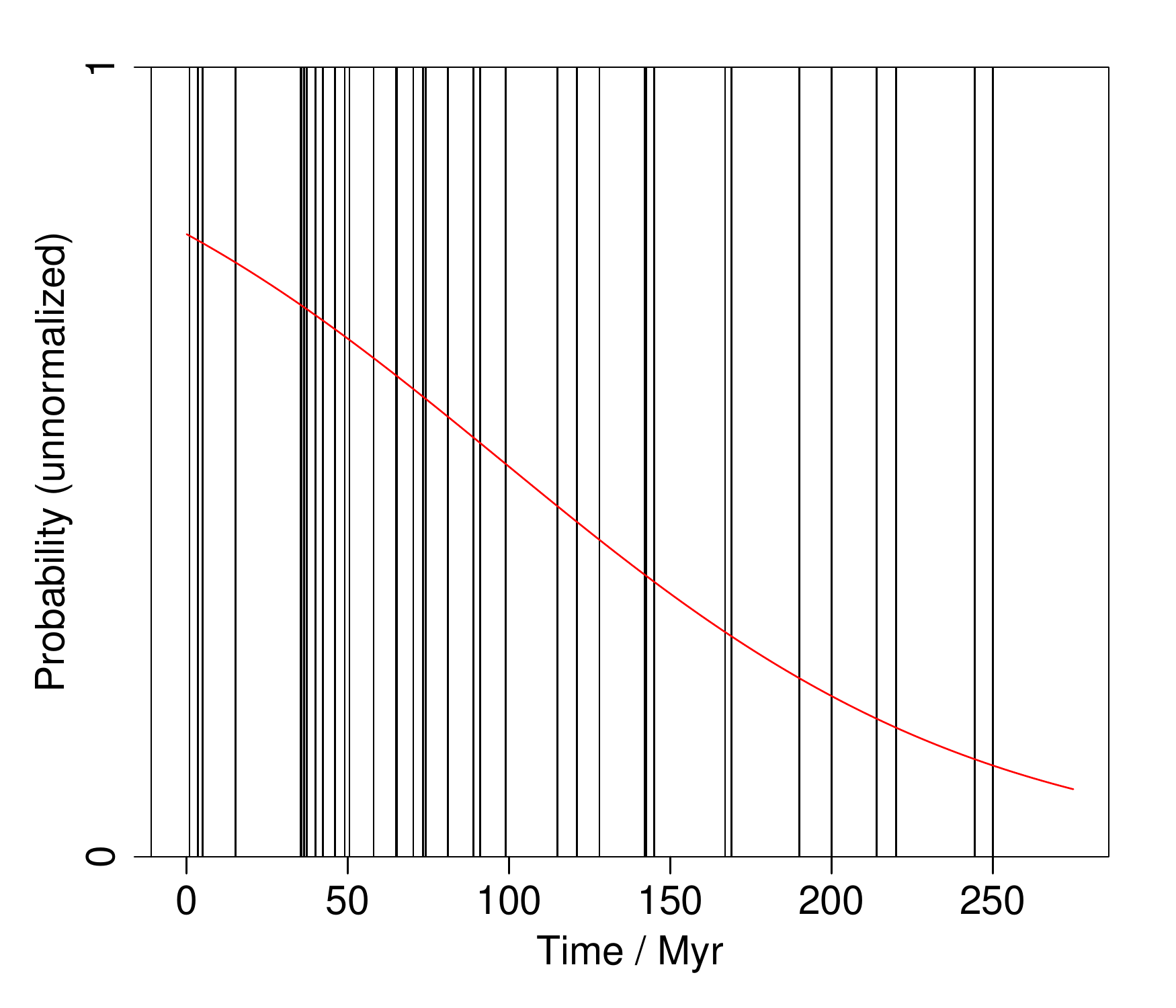}
\caption{The maximum likelihood solution of {\em SigProb} (red curve) for the basic250 data set (black lines). It has $\lambda,t_0=(-75,98)$
and a likelihood of $6.04\times10^{-101}$.
\label{fig:eidrun56_MLsolution}}
\end{center}
\end{figure}

Not only is {\em SigProb:Neg} the favoured model, it also has a large likelihood over a wide range of $t_0$ and $\lambda$. In other words, the evidence is not very sensitive to the exact parameter settings (nor to the prior).  This can be seen in Fig.~\ref{fig:eidrun56_loglikemap_MLpeak}: half of the parameter space below $\lambda=0$ has a likelihood within a factor of 10 of the maximum. (The model has a large Occam factor: see section~\ref{sect:discussion}) The maximum likelihood solution is plotted over the data in Fig.~\ref{fig:eidrun56_MLsolution}. Because the likelihood peak is so broad, we should not (need not) attribute much significance to this specific solution.

Is there evidence for a periodicity on top of this trend? We examine this using the four-parameter model {\em SinSigProb}
using the same parameter range adopted for the periodic and trend models (namely
all phases, periods of 10--125\,Myr, $-100 \leq \lambda <0$ and $0 < t_0 < 275$).
The evidence is $6.94\times10^{-102}$, a Bayes factor with respect to {\em SigProb:Neg} of just 0.40.  
If we reduce the parameter range, e.g.\ to 10-50\,Myr, $\lambda <25$ and $t_0 < 200$, then the evidence hardly changes.
We could possibly identify narrow, isolated regions of parameter space where {\em SinSigProb} is more probable, but this has no justification. 


In summary of the basic250 data set, we have found significant evidence for a trend of decreasing probability of cratering with lookback time relative to both constant and periodic models.  This is about the simplest model we can conceive after the constant one, and it also has more evidence than a model with a periodicity superimposed on either a trend or a constant background probability.



\subsection{Data set: ext250}

If I add the four craters for which I assigned age uncertainties, the evidence for the trend model relative to both the uniform model and the periodic models increases significantly (see Table~\ref{ressum}; the parameters for {\em SinSigProb} are the same as used for basic250).  This is not surprising because the additional craters are all comparatively young (1.07, 39, 46.5, 49\,Myr).

\subsection{Data set: full250}

I now add the 13 craters with upper age limits, noting that 12 of them are below 150\,Myr.
As they are {\em upper} age limits, it is not surprising that their inclusion increases the evidence for {\em SigProb:Neg} further. But the increase is enormous: the evidence is $1.53\times10^{-138}$ compared to $3.27\times10^{-145}$ for {\em UniProb}, a Bayes factor of $4.7\times10^{6}$.  

The evidence for {\em SinProb10:125} is very small,
giving a negligible Bayes factor relative to {\em SigProb:Neg}. The highest peak in the periodogram is still around 35\,Myr, but its evidence is $10^5$ times smaller than the evidence for {\em SigProb:Neg}.
Modelling the data using a periodicity on top of a trend -- with {\em SinSigProb} -- increases the evidence for proper periods (i.e.\ below 125\,Myr) dramatically over using just {\em SinProb}. But they are still insignificant compared to a pure trend. For example, the evidence for {\em SinSigProb} over periods of 10--50\,Myr and $\lambda < 0$ is $7.97\times10^{-140}$, a Bayes factor of 0.052 relative to {\em SigProb:Neg}. All peaks in the periodogram are also insignificant. The data are better described with a pure trend.


In summary, adding the 13 craters with only upper age limits radically increases the evidence for a negative trend, and radically decreases the evidence for either a periodicity or a periodicity plus negative trend, relative to the simple negative trend.

\subsection{Data set: large400}

The trend detected in the previous data sets might reflect a preservation bias in the geological record.
When extending the analysis to older craters we can try to avoid this bias by only including larger craters. The large400 data set
comprises 18 craters larger than 35\,km with ages up to 400\,Myr. 
The Bayes factor BF({\em SinProb10:400}$/${\em UniProb}) is 0.27. 
We see several low peaks in the periodogram, the four highest being 34\,Myr (BF\,=\,6.5), 18\,Myr and 13.5\,Myr (both BF\,=\,5.5), and 100\,Myr (BF\,=\,4). 
As there is no prior reason to expect a period at any of these, we cannot simply select one and claim it as evidence (albeit marginal) for that period, especially given that there are so many low peaks (and only 18 data points). 
Again we must look at the overall evidence for periodicity. For the period range 10--50\,Myr the Bayes factor
BF({\em SinProb10:50}$/${\em UniProb}) is 0.71, implying that the periodic and uniform models describe the data equally well.
But the periodic model with two free parameters for a pre-defined period range is arguably much less plausible a priori. 

There is also no evidence for a positive or negative trend in the data, with Bayes factors below 0.4 for the {\em SigProb} models (see Table~\ref{ressum}). 

In summary, the large400 data set is most plausibly described by {\em UniProb}.
Let us now turn to a discussion of the complete set of results as well as the method and principles on which it is based.

\section{Discussion}\label{sect:discussion}

I have found significant evidence for a decrease in the cratering rate with lookback time (model {\em SigProb:Neg}) over the past 250\,Myr for $d>5$\,km craters relative both to periodic models ({\em SinProb}, {\em SinBkgProb}, {\em SinSigProb}) and to a model with constant rate ({\em UniProb}).  As there is no strong evidence for a trend in the past 150\,Myr, this must come about primarily from a comparative lack of events between 150 and 250\,Myr BP. This may now seem obvious from  Figs.~\ref{fig:agediam_1} and ~\ref{fig:agediam_2}, but the analysis quantifies this and has also taken into account the age uncertainties.  No such trend is found for larger craters (d\,$>$\,35\,km) over the past 400\,Myr.  

These results could be explained by a decreasing probability of preservation/discovery for older craters of size 5--35\,km.
However, studies of lunar cratering suggest that the cratering rate during the past 500\,Myr was about twice as high as the average over the past 3.3\,Gyr (e.g.\ Shoemaker 1983). More immmediately relevant is the study of McEwen et al.\ (1997), who concluded that the cratering rate has increased up to the present by a factor of two during the past 300\,Myr. If correct, and if the Earth is assumed to have experienced the same bombardment history, then this 
is consistent with my inferred increase of impact probability from 250\,Myr BP up to the present. The single most probable solution for the trend model is shown in Fig.~\ref{fig:eidrun56_MLsolution}.

These conclusions are obviously based on the data we currently have. It is quite possible that a significant revision of the ages or the age uncertainties, or the inevitable discovery of more craters in the future, will lead us different conclusions based on the same analysis. More craters may permit a better distinction between more complex models.

I will now discuss some aspects of the method, and compare the present analysis with previous work.

\paragraph*{Significance assessment.} The significance of a model can only be assessed relative to the significance of some other model. There is no absolute. In frequentist statistics one normally selects some ``noise'' or ``background'' model against which to compare a statistic measured on the real data. For example, with the classical periodogram the significance is usually determined from the distribution of the power achieved by a noise model. This may indicate that the periodic model is the better of the two, but both might be bad: there may be a third model which is better still. 
We saw an example of this in Fig.~\ref{fig:eidrun43_SinProbmodels}, where the bottom left-hand panel is the best-fit truly periodic solution.
It was significant relative to {\em UniProb}, but insignificant relative to {\em SigProb:Neg}.


\paragraph*{Why we should not rely solely on periodogram peaks.} As has already been demonstrated in section~\ref{sect:simulations}, reliance on observing a peak in the periodogram -- even when normalized to the true model -- often results in erroneously claiming the periodic model to be a better explanation than the true one. The reason is that the periodogram has one free parameter (period), and we can sometimes find a specific value of this parameter which produces a better fit than the simpler uniform model (which has no free parameters). A model with even more free parameters may fit better still. But a model with {\em fitted} parameters is a priori less plausible than a model with no fitted parameters.  Unless we have independent information to assign the model parameters, we cannot fit them and {\em then} compare that model on an equal footing with a model which has not been fit. Instead we must compare models ``as a whole'' (e.g.\ over some period range). 
We saw an example of this in section~\ref{sect:sim_trend}.

\paragraph*{Occam factor.} The conclusion of the previous discussion is not that more complex models are always penalized. They are not. What counts is how the plausibility of the model is changed in light of the data. This can be understood by the concept of the {\em Occam factor}. If the likelihood function is dominated by a single peak, then we can approximate the evidence (equation~\ref{eqn:evidence}) with
\begin{equation}
\underbrace{P(D|M)}_\text{Evidence} \,=\, \underbrace{\mathcal{L}(\hat{\theta})}_\text{best fit likelihood} \times\;  \underbrace{\frac{\Delta \theta_{\rm posterior}}{\Delta \theta_{\rm prior}}}_\text{Occam factor}
\end{equation}
where $\mathcal{L}(\hat{\theta})$ is the likelihood at the best fit solution, $\Delta \theta_{\rm prior}$ is the prior parameter range and $\Delta \theta_{\rm posterior}$ is the posterior parameter range (the width of the likelihood peak) (see, e.g.\ MacKay 2003).  The Occam factor (which is always less than or equal to one) measures the amount by which the plausible parameter volume shrinks on account of the data. For given $\mathcal{L}(\hat{\theta})$, 
a simple or general model will fit over a large part of the parameter space,
so $\Delta \theta_{\rm prior} \sim \Delta \theta_{\rm posterior}$ and the Occam factor is not significantly less than one.  We saw an example of this in Fig.~\ref{fig:eidrun56_loglikemap_MLpeak}.  In contrast, a more complex model, or one which has to be more finely tuned to fit the data, will have a larger shrinkage, so $\Delta \theta_{\rm posterior} \ll \Delta \theta_{\rm prior}$. We saw this for the periodic models at short periods (e.g.~Figs.~\ref{fig:eidrun43_SinProbmodels} and~\ref{fig:eidrun43_periodogram}), in which only a very specific period was a good fit to the data. In this case the Occam factor is small and the evidence is reduced. Of course, if the fit is good enough then $\mathcal{L}(\hat{\theta})$ will be large, perhaps large enough to dominate the Occam factor and to give the model a large evidence. We saw this with the simulated periodic time series for the {\em SinProb} model (Fig.~\ref{fig:testrun1003_periodogram}).

This concept helps us to understand how the Bayesian approach accommodates model complexity, something generally lacking in frequentist approaches.  If we assess a model's evidence only by looking at the maximum likelihood solution (or the maximum over one parameter, the period), then we artificially compress the prior parameter range, increasing the Occam factor.

\paragraph*{Parameter prior distributions.} 
As the model evidence is the likelihood averaged over the prior parameter range (for uniform priors), this raises the issue of what this range should be. This is often the main perceived difficulty with Bayesian model comparison, and for some people this dependence on prior considerations is undesirable. Yet it is both logical and fundamentally unavoidable, because Bayesian or not, the prior parameter range is an intrinsic part of the model. Changing the parameter range changes the model, so will change the evidence.  {\em SinProb10:50} is totally different from {\em SinProb100:150}, for example. If we are comfortable with deciding which are the plausible models to test, we must also be willing to decide what are the plausible parameter ranges to test.  To some extent we can be guided by the context of the problem and the general properties of the data or the experiment, such as the sensitivity limits. For periodic models it seems obvious that we should use the whole phase range and that we should not include at ``periods'' much larger than the duration of observations (as these are more like trends). For {\em SigProb} I have actually used a rather broad range of its two parameters, even though some of this parameter space is a priori implausible, e.g.\ $\lambda=0$ gives a probability of zero to one side of $t_0=0$.

More generally, the evidence is the likelihood averaged over the parameter prior distribution. There are often cases where we would not want to use a uniform distribution.  It can be difficult to choose the ``correct'' prior distribution, and this choice may affect the results.  Yet whether we like it or not, interpreting data is a subjective business: Just as we choose which experiments to perform, which data to ignore, and which models to test, so we must decide what model parameters are plausible.  This seems preferable to ignoring prior knowledge or, worse, to pretending we are not using it.

In general, a probability density function is not invariant with respect to a nonlinear transformation of its parameters.  As already discussed in section~\ref{sect:priorchoice}, I could equally well have used frequency rather than period to calculate the evidence for periodic models: there is no ``natural'' parameter here. This would not change the error model of the data (eqn.~\ref{eqn:gaussage}), and the value of $P(t_j | \theta, M)$ at period $T$ is the same as when calculated at frequency $1/T$, so the likelihoods are unchanged. But as the model evidence is the average of the likelihoods over the prior, then the evidence would change if we adopted a prior which is uniform over frequency rather than over period. Thus the choice of parametrization becomes one of choice of prior. As neither parametrization is more natural than the other -- a prior uniform over frequency does not seem to be more correct than one uniform over period -- this remains a somewhat arbitrary choice. For this reason I repeated all of the analyses using periodic models with a prior uniform in frequency. Sometimes the evidence was slightly higher, sometimes lower, but the significance of the Bayes factors was not altered. The conclusions are robust to this change of prior/parametrization.

\paragraph*{Model priors.}  I have used Bayes factors to compare pairs of models. Models are treated equally, so a significant deviation from unity gives evidence for one model over the other.  However, if the models have different complexities (or rather, different prior plausibilities) and the Bayes factor is unity, then rather than being equivocal we may tend to prefer the simpler model. This is because we normally give the less plausible model a lower model prior probability (equation~\ref{eqn:modelpost2}). We can include these priors by reporting instead the {\em posterior odds} 
\begin{equation} 
\frac{P(D|M_k)}{P(D|M_0)} \, \frac{P(M_k)}{P(M_0)} \,=\, BF_{k0} \frac{P(M_k)}{P(M_0)}\ .  
\end{equation}
It is not obvious that all of the models I have considered should have equal model priors. 
For example, {\em SinSigProb} is arguably less plausible than {\em SigProb} a priori.

\paragraph*{Posterior probabilities and plausible models.} Ideally we would calculate model posterior probabilities, $P(M|D)$. This can only be done if we calculate the evidence for all {\em plausible} models, those with $P(M)$ which is not vanishingly small. In the current problem I have conceivably included most of the plausible parameter space for the stated models:
These non-deterministic models are quite flexible in describing general shapes. For example, I found that a model with periodic Gaussians (with three parameters) looked very similar to {\em SinProb}.  (Given sufficient physical reason, we could of course define more complex models, e.g.\ with a variable period or amplitude.)  Assuming that {\em SigProb:Neg}, {\em SigProb:Pos}, {\em UniProb} and {\em SinProb10:125} are the only plausible models, and assigning them equal priors, then for the basic250 data set, the posterior probability of {\em SigProb:Neg} from equation~\ref{eqn:modelpost3} is $(1 + 8.50\!\times\!10^{-6} + 5.95\!\times\!10^{-3} + 6.65\!\times\!10^{-4} )^{-1} = 0.993$.


\paragraph*{Some problems with frequentist hypothesis testing.} 
The motivation for the current work was to apply rigorous methodology to modelling impact crater time series, overcoming
some limitations of frequentist hypothesis testing (for a discussion of these problems see, e.g., Berger \& Sellke 1987, Marden 2000, Berger 2003, Jaynes 2003, Christensen 2005, Bailer-Jones 2009)
To summarize, the main problems which can occur are as follows.
\begin{enumerate}
\item Failure to take into account all plausible models. 
We see from equation~\ref{eqn:modelpost3} that the model posterior probability increases monotonically as
the sum (over the alternative models) decreases.
If we neglect plausible alternative models, the sum is smaller than it should be and the posterior probability of $M_0$ is artificially increased. This issue applies to any analysis, Bayesian or otherwise.
\item Incorrectly estimating significance by comparison to an inappropriate model (the ``null'' hypothesis).
\item Reliance on maximum likelihood or periodogram peak solutions without any regard to model plausibility/complexity (potentially leading to overfitting).
\item Failure to actually test the model of interest. A null hypothesis ($M_0$) is rejected and this is assumed to imply acceptance of the model of interest. This is only possible when there are only two plausible models which are mutually exclusive and exhaustive (rare in the physical sciences). Otherwise all models must be explicitly tested.
\item Use of a statistic to reject a model if that statistic is {\em more extreme} than the value given by the data. This is the usual approach taken with p-values for example, and is used in most periodograms, including the Lomb--Scargle. The reference to values not observed cannot be justified, but the methods are forced to on account of the previous point (failure to actually calculate an evidence for the model of interest);
\item Incorrect interpretation of p-values.
Frequentist statistics interprets a small value of $P(D|M_0)$ to be evidence for the alternative hypothesis, $M_1$. But that is given by $P(M_1|D)$, which we can {\em only} calculate if we also know $P(D|M_1)$ (equation~\ref{eqn:modelpost2}).
An example illustrates the difference. Suppose $P(D|M_0)=0.01$. This is interpreted as evidence against the null hypothesis $M_0$ at p\,$=$\,0.01.
But if the evidence for the alternative hypothesis is higher, but still relatively low, e.g.\
$P(D|M_1)=0.05$, then we see that 
$P(M_0|D) = 1/(1 + 0.05/0.01) = 0.17$ (for equal model priors). This is lower than $P(M_1 | D)$, but is not low enough to ``rule out $M_0$ at the 1\% level''. P-values frequently overestimate the significance.
\end{enumerate}
This is not to say that frequentist hypothesis testing and p-values are worthless. 
If we only have one model (e.g.\ ``Gaussian random noise''), then it may be hard to define an explicit model for its complement, in which case Bayesian model comparison is awkward. 
Here a low p-value is a useful indication that this model may not be a good explanation, prompting the search for alternatives.

\paragraph*{Previous studies of cratering periodicity.}  
Using similar data to those used here, several studies have claimed evidence for periodicities in the impact crater data, whereas several others conclude no evidence for periodicity. (Some studies are quite old, when fewer impact craters had been discovered or dated). 
A non-exhaustive list of the studies and their main results follows. 
\begin{itemize}
\item Alvarez \& Muller (1984): period at 28.4\,Myr for 11 craters with  $t<250$\,Myr, $\sigma_t<20$\,Myr, $d>5$\,km using a classical periodogram method. Jetsu \& Pelt (2000) claimed that this period (and potentially other claims of periodicity) is a artefact caused by rounding ages.
\item Rampino \& Stothers (1984): period at  31\,Myr for 41 craters with $t<250$\,Myr using a correlation method. 
This work was strongly criticised by Stigler (1985).
\item Grieve et al.\ (1985): periods found at 13.5, 18.5, 21 and 29\,Myr for a set of 26 craters with $t<250$\,Myr, $d=$\,5--10\,km using the method of Broadbent (1956). This method essentially defines a statistic which measures the deviation of a set of events from strict periodicity, and then estimates the probability that a uniform distribution would produce this value of the statistic (or smaller).
No periodicity was found for $d>10$\,km.
Partly because the extracted period depends on the data subset used, Grieve et al.\ concluded a lack of evidence for any true periodicity.
\item Grieve et al.\ (1988): doubt cast on previously claimed periods around 30\,Myr on a set of 27 craters, noting also that the correct period often cannot be found with the Broadbent method when there is a superimposed uniform random component. They found weak evidence for periods at 16\,Myr and 18--20\,Myr, the latter predominantly due to 10 craters in the past 40\,Myr. (See also Grieve 1991.)
\item Yabushita (1991): periods found at 16.5, 20 and 50 \,Myr for smaller craters (d\,$<$\,10\,km) in some data sets using the statistic of Broadbent, but different conclusions are reached depending on what significance test is adopted. He ultimately concludes that it is premature to claim evidence for periodicity.
\item Yabushita (1996): a period at 30\,Myr is found after examining various data sets (again using the Broadbent method), but it is found to be insignificant when a trend component (exponential decay) is included in the modelling.
\item Montanari et al.\ (1998): no periodicity found for 33 craters with $t<150$\,Myr, $\sigma_t<5$\,Myr, $d>5$\,km using a clustering method which looks at the (uncertainty weighted) age differences between craters.
\item Napier (1998): period at 13.4\,Myr, suggested to be a harmonic of a $27\pm1$\,Myr, from a set of 28 craters with $t<250$\,Myr, $\sigma_t<10$\,Myr, using a classical periodogram.
\item Stothers (1998): period at 36\,$\pm$\,1 Myr (using a variation of the Broadbent method), but concluded to be insignificant. (He also criticises some earlier period searching work.)
\item Yabushita (2004): period at 37.5\,Myr from a set of 91 craters with $t<400$\,Myr (including craters which have upper age limits provided that limit is $t<1$\,Myr) using a Lomb--Scargle periodogram in which the crater size or impact energy is taken as the dependent variable. The analysis yields a p-value for periodicity between 0.02 and 0.1 (depending on which craters are used). 
\item Chang \& Moon (2005): period at 26\,Myr for various subsets with $t<250$\,Myr, $d>5$\,km using the Lomb--Scargle periodogram.
\item Napier (2006): ``weak'' periods at 24, 35 and 42\,Myr (depending on what we consider to be harmonics) in a set of 
40 craters with $t<250$\,Myr, $\sigma_t<10$\,Myr, $d>3$\,km using a clustering method which examines the number of nearest neighbours.
\end{itemize}
Some studies have combined impact data with mass extinction time series (e.g.\ Napier 1998) or tried to identify correlations between them (e.g.\ Matsumoto \& Kubotani 1996).  Reviews and discussions can be found in Grieve (1991) and Grieve \& Pesonen (1996).  Numerical simulations of what signals can be detected in these kinds of data sets were presented in Heisler \& Tremaine (1989) and Lyytinen et al.\ (2009).

The above summary makes clear that a large number of periods have been claimed, some of which can be identified as potential harmonics of others.  Several of these periods, e.g. 11.5, 13.5, 18, 35\,Myr, I also see as a local maximum in my likelihood distributions or periodograms. But in no case do I find any of them to be significant.

Most of the above studies employ frequentist hypothesis testing and suffer from one or more of the problems outlined previously. In particular, many compare the value of some statistic measured on the data to the values obtained by a noisy ``reference'' model. This is typically a uniform random distribution (akin to {\em UniProb}).
Only the study of Yabushita (1996) examined a trend component in the analysis, and when this was included there was insufficient evidence for periodicity.  As already demonstrated and discussed, {\em if the reference model is inappropriate and if other plausible models are ignored, then the significance of any periodicity is overestimated}.  Moreover, by only considering the evidence in a single period we overfit the model, leading to a claim of periodicity where none exists.  Note also that focusing on a single period (or narrow range) because it has been found in a previous study does not give an independent claim for periodicity, because we only have one crater record. This would amount to reusing the data, thereby increasing the evidence for the period artificially.


There may be a human desire to find and report periods. Several studies give some prominence to detected periods, but draw less attention to their limited statistical significance (e.g.\ Yabushita 1991, Stothers 1998).
The motivation to do this may be a lack of confidence in the robustness of the test, anticipation that a later analysis will confirm the period(s) with significance, or because the period is close to another period (itself possibly insignificant) found in other studies, in biodiversity data, or in models of solar motion (e.g.\ Stothers 1998).  But significance is everything: non-periodic models can give rise to superficially significant periods (section~\ref{sect:sim_trend}; see also Stigler \& Wagner 1987).  The (lack of) evidence for a periodicity in geological data or for an expected periodicity in astronomical phenomena is reviewed in Bailer-Jones~\cite{cbj09}.


\paragraph*{Summary of the main features of the method.}  

The analysis method developed in this article is quite general and is not limited to analysis of impact crater time series. Its main features are
\begin{itemize}
\item operation on time-of-arrival data
\item description of time series as stochastic models (more appropriate to the impact phenomena than deterministic models)
\item consistent use of age uncertainties (obviating the need to remove ``poor'' data)
\item ability to include craters with upper age limits (censored data) consistently
\item use of Bayesian evidence calculation. Avoidance of p-values or ad hoc statistics
\item comparison of multiple models (rather than relying on a single ``null'' hypothesis)
\item use of proper parameter prior distributions, which are considered as an intrinsic part of the model.
\end{itemize}
The method should not be very sensitive to age errors provided the age uncertainties are approximately correct, although detailed, systematic testing of this has not yet been performed. We may also want to see how robust the conclusions are to the inclusion/removal of craters around the 5\,km diameter limit. On the other hand, we have seen that the addition/removal of a few craters does not change the conclusions, as we would expect.

\section{Conclusions}\label{sect:conclusions}

I find no evidence for a periodic variation in the impact crater rate the past 150\,Myr or 250\,Myr for craters with diameter above 5\,km, relative to two other plausible -- but quite broad -- models: constant and monotonically varying probability with time. Compared to the uniform model, there is significant evidence for a monotonic decrease in the impact rate with look back time over the past 250\,Myr, but not over the past 150\,Myr. However, introducing craters with upper age limits into the analysis does give significant evidence for this trend even within the past 150\,Myr.  The physical interpretation is either an intrinsic variation in the impact probability or a variation in the crater preservation/discovery probability (i.e.\ we are less likely to find older craters).  The former is consistent with some studies of lunar cratering.  For very large craters ($d>35$\,km) over the past 400\,Myr the best fitting model is the constant probability one, which is consistent with no preservation/discovery bias for such large craters. Given what we know about crater erosion and infilling, the preservation/discovery bias is a plausible explanation (e.g.\ Grieve 1991).  It remains possible that there is a periodic variation on top of this. I find no evidence for this, although such a complex signal would be difficult to distinguish.  Further simulations are necessary to explore what other kinds of signal could be reliably detected in these geological data.


Contrary to claims made in the literature, we can draw useful conclusions from events with large or variable age uncertainties, provided we use these uncertainties correctly. Larger uncertainties may make it harder to distinguish between models, but they do not invalidate the concept of model comparison. 

Other studies claiming periodicity have probably overestimated the significance of the detected periods. This can occur if the significance is assessed relative to a null hypothesis of a model which is poorly supported by the data (here the uniform model), rather than to other non-periodic models which may be much better supported (here the trend model). The wrong conclusion can also be reached if we rely on a periodogram peak or the maximum likelihood solution. I have shown via simulations how these can often claim a periodicity where none exists. The reason is that they fail to assess the evidence for the model as a whole, examining only the likelihood of an ``overfit'' solution. 


\section*{Acknowledgements}

I would like to thank Rainer Klement, Chao Liu, Kester Smith and Vivi Tsalmantza for fruitful discussions during the development of this work, as well as David Hogg and the referee for thoughtful comments on the manuscript.  I am grateful to the staff of the Planetary and Space Science Centre at the University of New Brunswick, Canada, for their work in compiling and maintaining the Earth Impact Database, and to those who have contributed their data to it.


\appendix
\section{Models with a uniform frequency prior}\label{sect:appendix}

The periodic models described in section~\ref{sect:tsmodels} have been re-run using frequency instead of period as the model parameter, and with a uniform prior over frequency.  As discussed in the main paper, it turns out that this has no relevant impact on the results, i.e.\ the conclusions are robust with respect to this reparametrization or change of prior. So in the interests of brevity the results are show here for just one model and data set.

\begin{figure} 
\begin{center}
\includegraphics[width=0.5\textwidth, angle=0]{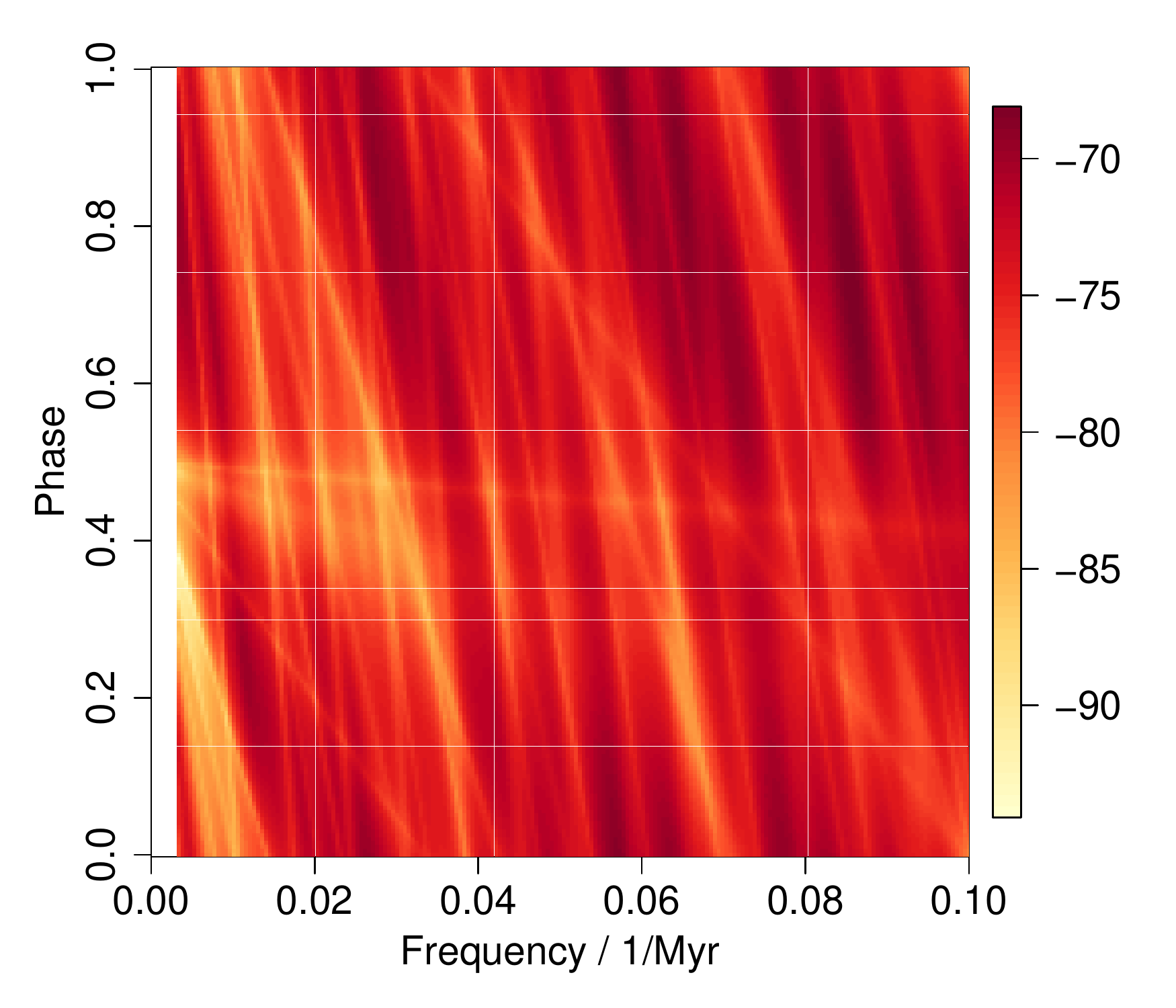}
\caption{Log likelihood distribution as a function of frequency and phase for the {\em SinProb} model with a prior uniform over frequency for the basic150 data set
\label{fig:eidrun72_loglikemap}}
\end{center}
\end{figure}

\begin{figure} 
\begin{center}
\includegraphics[width=0.40\textwidth, angle=0]{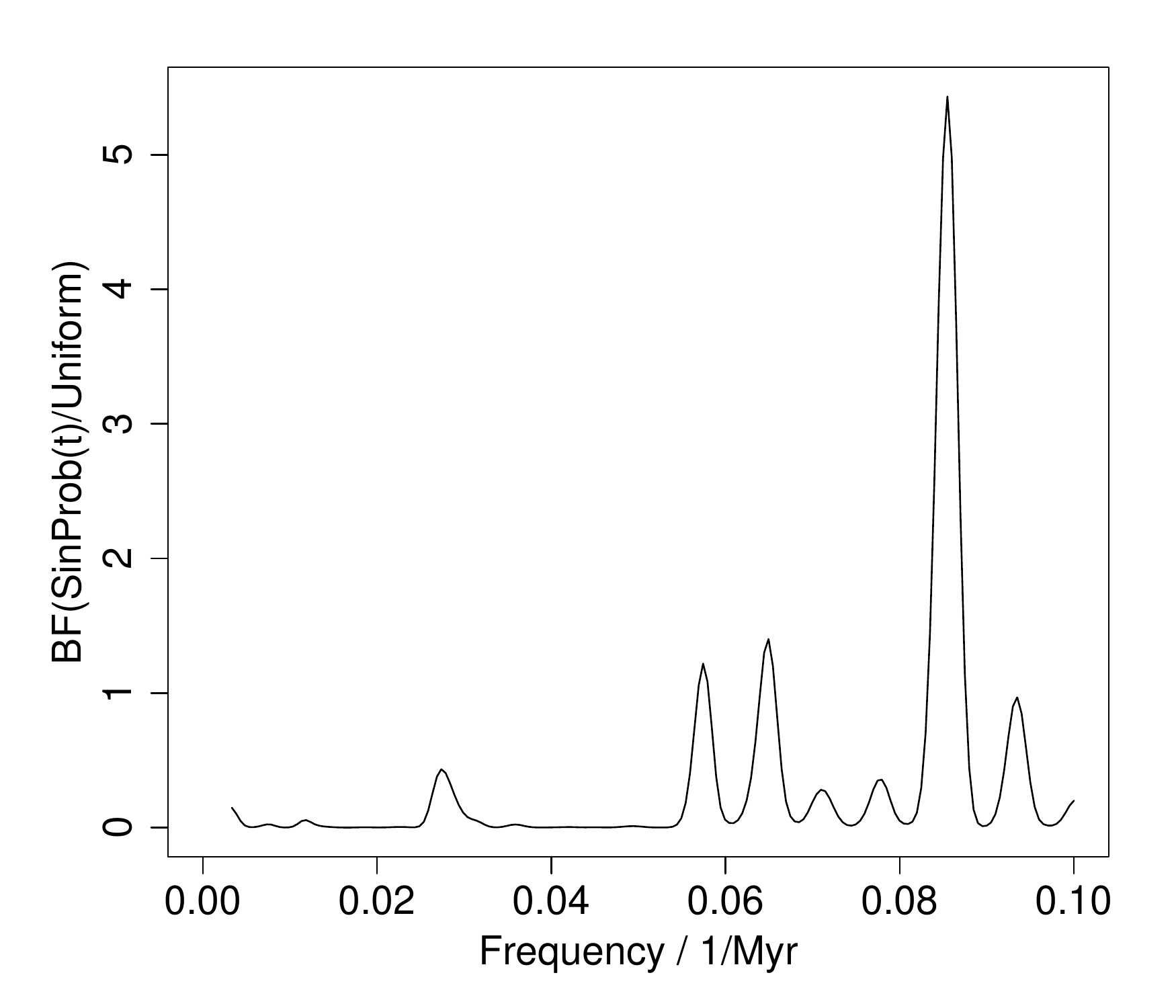}
\caption{Bayesian periodogram for the {\em SinProb} model with a prior uniform over frequency, for the basic150 data set
\label{fig:eidrun72_periodogram}}
\end{center}
\end{figure}

Fig.~\ref{fig:eidrun72_loglikemap} shows the likelihood distribution for the {\em SinProb} model applied to the basic150 data set. The corresponding periodogram, formed by marginalizing over the phase, is show in Fig.~\ref{fig:eidrun72_periodogram}. These two figures may be compared to the two corresponding figures for
the model with the prior uniform over period, Figs.~\ref{fig:eidrun42_loglikemap} and \ref{fig:eidrun42_periodogram}.
The two periodograms actually show peaks at identical periods. Indeed, if we replot Fig.~\ref{fig:eidrun72_periodogram} in terms of period (i.e.\ transform the abcissa), we get a plot almost identical to Fig.~\ref{fig:eidrun42_periodogram}, but with slightly higher values on the ordinate. The higher values is simply a result of the smoothing scale used to produce the plot: we get these slightly higher values in Fig.~\ref{fig:eidrun42_periodogram} too if we use a smaller smoothing scale.
(This also reflects the fact that a fixed smoothing scale parameter in frequency corresponds to a variable one in period, and vice versa.)
The maximum likelihood solution is exactly the same.
The evidence for common period ranges is actually slightly different, e.g.\ $6.01\times10^{-71}$ for periods of 10--50\,Myr 
with the uniform frequency prior (Bayes factor of 0.37 with respect to {\em UniProb}), 
compared to $2.67\times10^{-71}$ before. This reflects the different ``weighting'' of the likelhoods that comes with using a different prior (see equation~\ref{eqn:evidence}). However, these changes are comparatively small, and do not alter the conclusions. In many other cases the differences are smaller.

\begin{thebibliography}{99}


\bibitem[2003]{alvarez_2003}
Alvarez W., 2003, Astrobiology 3, 153
\vspace*{0.21em}

\bibitem[1980]{alvarez_etal_1980}
Alvarez L.W., Alvarez W., Asaro F., Michel H.V., 1980, Science 208, 1095
\vspace*{0.21em}

\bibitem[1984]{alvarez_muller_1984}
Alvarez W., Muller R.A., 1984, Nature 308, 718
\vspace*{0.21em}

\bibitem[2009]{cbj09} 
Bailer-Jones C.A.L., International Journal of Astrobiology 8, 213

\bibitem[1985]{bahcall_bahcall_1985} 
Bahcall J.N., Bahcall S., 1985, Nature 316, 706
\vspace*{0.21em}

\bibitem[2003]{berger_2003}
Berger J.O., 2003, Statistical Science 18, 1
\vspace*{0.21em}

\bibitem[1987]{berger_sellke_1987}
Berger J.O., Sellke T., 1987, Journal of the Americal Statistical Association 82, 112
\vspace*{0.21em}


\bibitem[]{} 
Broadbent S.R., 1956, Biometrika 43, 32
\vspace*{0.21em}


\bibitem[]{} 
Chang H.-Y., Moon H.-K.\, 2005, PASJ 57, 487
\vspace*{0.21em}

\bibitem[2005]{christensen_2005}
Christensen R., 2005, The American Statistician 59, 121
\vspace*{0.21em}

\bibitem[1984]{davis_etal_1984}
Davis M., Hut P., Muller R.A., 1984, Nature 308, 715
\vspace*{0.21em}

\bibitem[]{} 
Deutsch A., Sch\"arer U., 1994, Meteoritics 29, 301
\vspace*{0.21em}

\bibitem[]{} 
Le Feuvre M., Wieczorek M.A., 2008, Icarus 197, 291
\vspace*{0.21em}

\bibitem[2005]{gregory_2005}
Gregory, P., 2005, {\em Bayesian logical data analysis for the physical sciences}, Cambridge University Press
\vspace*{0.21em}

\bibitem[]{} 	
Gardner E., Nurmi P., Flynn C., Mikkola S., 2011, MNRAS 411, 947
\vspace*{0.21em}

\bibitem[]{} 
Grieve R.A.F., 1991, Meteoritics 26, 175
\vspace*{0.21em}

\bibitem[]{} 
Grieve R.A.F., Pesonen L.J., 1996, Earth, Moon \& Planets 72, 357

\bibitem[1985]{grieve_etal_1985}
Grieve R.A.F., Sharpton V.L., Goodacre A.K., Garvin J.B., 1985, Earth and Planetary Science Letters 76, 1
\vspace*{0.21em}

\bibitem[1988]{grieve_etal_1988} 
Grieve R.A.F., Sharpton V.L., Rupert J.D., Goodacre A.K., 1988, Proceedings of the $18^{th}$ LPSC, 375
\vspace*{0.21em}

\bibitem[2004]{hallam_2004}
Hallam A., 2004, {\em Catastrophes and lesser calamaties. The causes of mass extinctions}, Oxford University Press
\vspace*{0.21em}

\bibitem[1989]{heisler_tremaine_1989}
Heisler J., Tremaine S., 1989, Icarus 77, 213
\vspace*{0.21em}

\bibitem[2003]{jaynes_2003}
Jaynes E.T., 2003, {\em Probability theory. The logic of science}, Cambridge University Press
\vspace*{0.21em}

\bibitem[]{} 
Jeffreys H., 2000, {\em Theory of Probability},  Cambridge University Press
\vspace*{0.21em}

\bibitem[2000]{jetsu_pelt_2000} 
Jetsu L., Pelt J., 2000, A\&A 353, 409
\vspace*{0.21em}

\bibitem[]{} 
Kass R.E., Raftery A.E., 1996, Journal of the Americal Statistical Association 90, 773
\vspace*{0.21em}

\bibitem[]{} 
Lyytinen J., Jetsu L., Kajatkari P., Porceddu S., 2009, A\&A 499, 601
\vspace*{0.21em}

\bibitem[]{} 
Mackay D.J.C., 2003, {\em Information Theory, Inference and Learning Algorithms}, Cambridge University Press
\vspace*{0.21em}

\bibitem[1996]{matsumoto_kubotani_1996}
Matsumoto M., Kubotani H., 1996, MNRAS 282, 1407
\vspace*{0.21em}

\bibitem[2000]{marden_2000}
Marden J.I., 2000, Journal of the Americal Statistical Association 95, 1316
\vspace*{0.21em}

\bibitem[]{} 
McEwen A.S., More J.M., Shoemaker E.M., 1997, J.\ Geophysical Research 102, 9231

\bibitem[]{} 
Montanari A., Campo Bagatin A., Farinella P., 1998, Planet.\ Space Sci.\ 46, 271
\vspace*{0.21em}

\bibitem[1998]{napier_1998}
Napier W.N., 1998, Celestial Mechanics and Dynamical Astronomy 69, 59
\vspace*{0.21em}

\bibitem[]{} 
Napier W.N., 2006, MNRAS 366, 977
\vspace*{0.21em}

\bibitem[1984]{rampino_stothers_1984}
Rampino M.R., Stothers R.B., 1984, Nature 308, 709
\vspace*{0.21em}

\bibitem[1983]{shoemaker_1983}
Shoemaker E.M., 1983, Ann.\ Rev.\ Earth Planet.\ Science 11, 461
\vspace*{0.21em}

\bibitem[1986]{shuter_klatt_1986} 
Shuter W.L.H., Klatt C., 1986,  ApJ 301, 471
\vspace*{0.21em}

\bibitem[1985]{stigler_1985}
Stigler S.M., 1985, Nature 313, 159
\vspace*{0.21em}

\bibitem[1987]{stigler_wagner_1987}
Stigler S.M., Wagner M.J., 1987, Science 238, 940
\vspace*{0.21em}

\bibitem[]{} 
Stothers 1998, MNRAS 300, 1098
\vspace*{0.21em}

\bibitem[1984]{torbett_smo_1984}
Torbett M.V., Smoluchowski R., 1984, Nature 311, 641
\vspace*{0.21em}

\bibitem[2008]{wick_napier_2008} 
Wickramasinghe J.T., Napier W.M., 2008, MNRAS 387, 153
\vspace*{0.21em}

\bibitem[]{} 
Yabushita S., 1991, MNRAS 250, 481
\vspace*{0.21em}

\bibitem[]{} 
Yabushita S., 1996, MNRAS 279, 727
\vspace*{0.21em}


\bibitem[]{} 
Yabushita S., 2004, MNRAS 355, 51
\vspace*{0.21em}

\end{thebibliography}
\end{document}